\documentclass[nohyperref]{article}

\usepackage{microtype}
\usepackage{graphicx}
\usepackage{subfigure}
\usepackage{booktabs} % for professional tables
\usepackage{xcolor}
\usepackage{setspace}
\usepackage{multirow}
\usepackage{multicol}
\usepackage{threeparttable}
\usepackage{tabulary}
\usepackage{colortbl}
\usepackage{ragged2e}
\usepackage{enumitem}
\usepackage{amsmath,bm}

\usepackage{hyperref}

\makeatletter
\DeclareFontFamily{U}{tipa}{}
\DeclareFontShape{U}{tipa}{m}{n}{<->tipa10}{}
\newcommand{\arc@char}{{\usefont{U}{tipa}{m}{n}\symbol{62}}}%
\newcommand{\arc}[1]{\mathpalette\arc@arc{#1}}
\newcommand{\arc@arc}[2]{%
  \sbox0{$\m@th#1#2$}%
  \vbox{
    \hbox{\resizebox{\wd0}{\height}{\arc@char}}
    \nointerlineskip
    \box0
  }%
}
\makeatother

\newcommand{\method}{ProtST}
\newcommand{\dataset}{ProtDescribe}

\setlist[itemize]{leftmargin=5mm}

\definecolor{temp}{RGB}{0,0,255}   
\definecolor{todo}{RGB}{255,0,0}
\definecolor{highlight}{RGB}{218,165,32}
\definecolor{decay}{RGB}{192,192,192}
\definecolor{r1}{RGB}{25,95,225}
\definecolor{r2}{RGB}{30,144,255}
\definecolor{r3}{RGB}{135,206,235}
\usepackage[accepted]{icml2023}

\icmltitlerunning{{\method}: Multi-Modality Learning of Protein Sequences and Biomedical Texts
}

\usepackage{enumitem}
\usepackage{xspace}
\usepackage{xcolor}
\usepackage{wrapfig}
\usepackage{caption}
\usepackage{amssymb}
\usepackage{multirow}
\usepackage{adjustbox}

\usepackage{thmtools}
\usepackage{thm-restate}

\newcommand{\mathbbm}[1]{\text{\usefont{U}{bbm}{m}{n}#1}}

\definecolor{myblue}{RGB}{68, 114, 196}
\definecolor{myorange}{RGB}{237, 125, 49}

\usepackage{amsmath}
\usepackage{bm}

\def\eqref#1{equation~\ref{#1}}
\def\1{\bm{1}}

\DeclareMathAlphabet{\mathsfit}{\encodingdefault}{\sfdefault}{m}{sl}
\SetMathAlphabet{\mathsfit}{bold}{\encodingdefault}{\sfdefault}{bx}{n}

\usepackage{cancel}

\begin{document}

\twocolumn[
%\icmltitle{Pretraining Structure-Based Encoders for Protein Representation Learning}
\icmltitle{{\method}: Multi-Modality Learning of Protein Sequences \\ and Biomedical Texts
% Enhancing Protein Sequence Pre-training and Understanding \\ by Biomedical Texts
}

\icmlsetsymbol{equal}{*}
\icmlsetsymbol{lead}{$\dagger$}

\begin{icmlauthorlist}
\icmlauthor{Minghao Xu}{equal,lead,mila,udem}
\icmlauthor{Xinyu Yuan}{equal,mila,udem}
\icmlauthor{Santiago Miret}{intel}
\icmlauthor{Jian Tang}{mila,hec,cifar}
\end{icmlauthorlist}

\icmlaffiliation{mila}{Mila - Qu\'{e}bec AI Institute}
\icmlaffiliation{intel}{Intel Labs}
\icmlaffiliation{udem}{Universit\'e de Montr\'eal}
\icmlaffiliation{hec}{HEC Montr\'{e}al}
\icmlaffiliation{cifar}{CIFAR AI Research Chair}

\icmlcorrespondingauthor{Minghao Xu}{minghao.xu@mila.quebec}
\icmlcorrespondingauthor{Santiago Miret}{santiago.miret@intel.com}
\icmlcorrespondingauthor{Jian Tang}{jian.tang@hec.ca}

\icmlkeywords{Protein Sequence Pre-training, Protein Sequence Understanding, Biomedical Text Understanding, Multimodal Representation Learning}

\vskip 0.3in
]



\printAffiliationsAndNotice{\icmlEqualContribution}
% \printAffiliationsAndNotice{}

%%%%%%%%%%%%%%%%%%%%%%%%%%%%%%%%%%%%%%%%%%%%%%%%%%%%%%%%%%%%

\begin{abstract}

% Existing protein language models (PLMs) learn protein representations mainly based on protein sequences, which can well capture the co-evolutionary information but cannot explicitly acquire protein functions, which is actually the end goal of protein representation learning. 
Current protein language models (PLMs) learn protein representations mainly based on their sequences, thereby well capturing co-evolutionary information, but they are unable to explicitly acquire protein functions, which is the end goal of protein representation learning. Fortunately, for many proteins, their textual property descriptions are available, where their various functions are also described. 
% Protein language models (PLMs) pre-trained on large-scale protein sequence corpus have achieved impressive results on a variety of protein understanding tasks. However, existing PLMs can hardly capture diverse biological and biomedical properties of proteins during pre-training, which limits their effectiveness on predicting various protein properties for downstream applications. 
% To tackle such a limitation, 
Motivated by this fact, we first build the \textbf{\dataset} dataset to augment protein sequences with text descriptions of their functions and other important properties. 
% which covers a broad range of protein properties. 
Based on this dataset, we propose the \textbf{\method} framework to enhance \textbf{Prot}ein \textbf{S}equence pre-training and understanding by biomedical \textbf{T}exts. During pre-training, we design three types of tasks, \emph{i.e.}, unimodal mask prediction, multimodal representation alignment and multimodal mask prediction, to enhance a PLM with protein property information with different granularities and, at the same time, preserve the PLM's original representation power. On downstream tasks, {\method} enables both supervised learning and zero-shot prediction. We verify the superiority of {\method}-induced PLMs over previous ones on diverse representation learning benchmarks. Under the zero-shot setting, we show the effectiveness of {\method} on zero-shot protein classification, and {\method} also enables functional protein retrieval from a large-scale database without any function annotation. 
% we verify the effectiveness of {\method} on protein classification and text-to-protein retrieval. 
% we verify the data efficiency of {\method}-induced zero-shot predictors and prove that these predictors can be further used to boost supervised learning models. 
% The source code will be released upon acceptance. 
Source code and model weights are available at \url{https://github.com/DeepGraphLearning/ProtST}.

\end{abstract}

%%%%%%%%%%%%%%%%%%%%%%%%%%%%%%%%%%%%%%%%%%%%%%%%%%%%%%%%%%%%

%%%%%%%%%%%%%%%%%%%%%%%%%%%%%%%%%%%%%%%%%%%%%%%%%%%%%%%%%%%%

\section{Introduction} \label{sec:intro}

Proteins serve as the mainstay governing diverse biological processes and life itself, inducing important applications in drug discovery~\cite{teague2003implications} and healthcare~\cite{world2007protein}. Recent studies have proven the great promise of machine learning methods in predicting protein structures~\cite{jumper2021highly,baek2021accurate} and functionality~\cite{meier2021language,gligorijevic2021structure}. Among these methods, protein language models (PLMs)~\cite{elnaggar2020prottrans,rives2021biological,lin2022language} pre-trained on large-scale protein sequence corpus succeed in acquiring powerful protein representations, which boost protein structure and function prediction~\cite{xu2022peer}.  
% which boost a broad range of downstream protein understanding tasks. 

Most existing PLMs~\citep{elnaggar2020prottrans,lu2020self,rives2021biological,lin2022language} learn protein representations based only on their sequences, which can well capture co-evolutionary information but cannot explicitly acquire protein functions and other important properties like their subcellular locations. Acquiring such function and property information is actually the end goal of protein representation learning. Fortunately, for many proteins, we can get access to their textual property descriptions in which their diverse functions are also described. This fact motivates us to study protein sequence representation learning enriched with diverse protein properties described by biomedical texts. 

To our best knowledge, OntoProtein~\citep{zhang2022ontoprotein} is the only existing PLM that explicitly captures protein properties. However, it learns a closed set of properties over a \emph{fixed biological knowledge graph} and thus can hardly generalize to unknown properties of new proteins. In comparison, by modeling \emph{textual protein property descriptions}, we can flexibly model the generalization from known properties to unknown ones based on the semantic correlation of their text descriptions, as shown by our zero-shot experiments (Secs.~\ref{sec:exp:zero} and \ref{sec:exp:t2p}). 

To attain biomedical-text-enhanced protein sequence representation learning, we first build the \textbf{\dataset} dataset, a paired dataset of protein sequences and textual property descriptions. We resort to the Swiss-Prot database~\cite{bairoch2000swiss} for high-quality protein annotations and construct each protein's property description with the selected annotations of it. 
% , forming a paired dataset of protein sequences and textual property descriptions, named as \textbf{\dataset}. 
{\dataset} incorporates the information of protein names, protein functions, subcellular locations and protein families, and these properties are described by biomedical texts with rich expressions. 

Based on this dataset, we propose the \textbf{\method} framework to enhance protein sequence pre-training and understanding by biomedical texts. 
% which benefits diverse types of downstream applications. 
During {\method} pre-training, to preserve the beneficial representation power of a conventional PLM on capturing co-evolutionary information, we adopt the \textbf{Unimodal Mask Prediction} task for masked protein modeling. On such basis, two multimodal pre-training tasks are designed to inject different granularities of pertinent protein property information into a PLM: \textbf{Multimodal Representation Alignment} injects integrated and general property information into the PLM, in which a biomedical language model is used to extract structured text representations of different property descriptions, and protein sequence representations are aligned to the corresponding text representations; \textbf{Multimodal Mask Prediction} models the fine-grained dependencies between residues in a protein sequence and property-descriptive words in its property description, in which a fusion module is employed to derive multimodal representations of residues and words, and, based on these fused multimodal representations, masked residues and words are predicted. 
% \textbf{Multimodal Representation Alignment} utilizes a biomedical language model to extract structured text representations of property descriptions and aligns protein sequence representations to the text representations, such that the PLM can acquire the integrated and general property information; \textbf{Multimodal Mask Prediction} employs a fusion module to derive multimodal representations of residues in a protein and words in its corresponding property description, and, based on such representations, masked residues and words are predicted to model the fine-grained dependencies between different residues and various property-descriptive words. 
For downstream applications, {\method} can conduct supervised learning with only the PLM and can also perform zero-shot prediction based on the aligned representation space of protein sequences and text descriptions. 
% During pre-training, the PLM is trained along with a biomedical language model (BLM) and a fusion module to solve three types of tasks: (1)~\emph{unimodal mask prediction} performs the standard masked protein modeling to preserve the beneficial representation power of a conventional PLM; (2)~\emph{multimodal representation alignment} utilizes the BLM to extract structured representations of property descriptions and aligns protein sequence representations to the text representations, such that the PLM can acquire the integrated and general property information; (3)~\emph{multimodal mask prediction} employs a fusion module to derive multimodal representations of residues in a protein and words in its corresponding property description, and, on such basis predicts masked residues and words to model the fine-grained dependencies between different residues and various property-descriptive words. By solving these three tasks simultaneously, the PLM can acquire pertinent protein property information with different granularities. For downstream applications, {\method} can conduct supervised learning with only the PLM and can also perform zero-shot prediction based on the aligned representation space of protein sequences and text descriptions. 

We investigate the PLMs trained under {\method} by representation learning and zero-shot prediction. 
% We demonstrate the representation learning and zero-shot prediction capabilities of the PLMs trained under {\method}. 
For representation learning, we verify their superior performance over previous masked language modeling and knowledge-enhanced PLMs on 11 standard benchmarks for protein localization prediction, fitness landscape prediction and protein function annotation (Sec.~\ref{sec:exp:repr}). 
%For zero-shot protein classification, we show the data efficiency of {\method}-induced zero-shot classifiers against various few-shot classifiers in Sec.~\ref{sec:exp:zero:eff}, and the zero-shot classifiers are also proven to be able to enhance the performance of supervised learning models via ensemble in Sec.~\ref{sec:exp:zero:enhance}. 
For zero-shot protein classification, {\method}-induced zero-shot classifiers show better data efficiency against various few-shot classifiers (Sec.~\ref{sec:exp:zero:eff}), and are proven to be able to enhance the performance of supervised learning models via ensemble (Sec.~\ref{sec:exp:zero:enhance}). 
For zero-shot text-to-protein retrieval, we verify the effectiveness of {\method} on 
retrieving functional proteins from a large-scale database without any function annotation (Sec.~\ref{sec:exp:t2p}). 
% For zero-shot prediction, we prove the data efficiency of the {\method}-induced zero-shot predictor against various few-shot predictors, and the zero-shot predictor is also verified to be able to enhance the performance of supervised learning models via ensemble. Given a large protein sequence corpus without any function annotation, the {\method}-induced PLM is proved to be able to accurately retrieve proteins with various specified functions. 
% In addition, we conduct extensive analytical studies to investigate different components of {\method}. 

\iffalse
\textcolor{temp}{
(Temp) To summarize, our contributions are as follows:
\begin{enumerate}
    \item We contribute the {\dataset} dataset, a collection of paired protein sequences and protein property descriptions, which covers a diverse set of protein properties. 
    \item We propose the {\method} framework to enhance a PLM with the protein property information with different granularities, and {\method} enables both supervised learning and zero-shot prediction on downstream tasks. 
    \item We demonstrate the effectiveness of {\method}-induced PLMs on both representation learning and zero-shot prediction, and we investigate different framework components with extensive analytical studies. 
\end{enumerate}
}
\fi

%%%%%%%%%%%%%%%%%%%%%%%%%%%%%%%%%%%%%%%%%%%%%%%%%%%%%%%%%%%%

%%%%%%%%%%%%%%%%%%%%%%%%%%%%%%%%%%%%%%%%%%%%%%%%%%%%%%%%%%%%

\section{Preliminaries} \label{sec:pre}

\subsection{Problem Definition} \label{sec:pre:def}

In the pre-training phase, we study the problem of learning informative protein sequence representations guided by the proteins' associated biomedical text descriptions. In this problem, a protein $P=(S, T)$ is represented by an amino acid sequence $S=[s_1, s_2, \cdots , s_n]$ with $n$ amino acids (\emph{a.k.a.}, residues) and a text description $T=[t_1, t_2, \cdots , t_m]$ with $m$ word tokens. 
% Each residue is from a vocabulary $\mathcal{V}_S$ composed of 20 common amino acids and an unknown token, and each word token is from a vocabulary $\mathcal{V}_T$ constructed based on biomedical text corpus. \
Given a pre-training dataset with $N$ proteins $\mathcal{P} = \{ P_1, P_2, \cdots , P_N \}$, our goal is to extract effective protein representations by fully utilizing the information from their sequences and descriptions. The extracted protein representations are expected to boost various downstream tasks by supervised learning or zero-shot prediction. 

%%%%%%%%%%%%%%%%%%%%%%%%%%%%%%%%%%%%%%%%%%%%%%%%%%%%%%%%%%%%

\subsection{Protein Language Models} \label{sec:pre:protein_lm}

Protein language models (PLMs)~\cite{elnaggar2020prottrans,rives2021biological,meier2021language,lin2022language} pre-trained on large-scale protein sequence corpus have shown impressive results on protein function~\cite{meier2021language} and structure~\cite{lin2022language} prediction. 
% diverse protein understanding tasks, including protein function prediction~\cite{meier2021language} and protein structure prediction~\cite{lin2022language}. 
PLMs are commonly trained by masked protein modeling, in which partial residues are masked at input and predicted based on the context. 
% their contextualized representations extracted by the model. 
In this work, we select three state-of-the-art PLMs, ProtBert~\cite{elnaggar2020prottrans}, ESM-1b~\cite{rives2021biological} and ESM-2~\cite{lin2022language}, as baselines and seek to enhance their representation power by 
% incorporating biomedical language modeling in addition to protein sequence modeling.
modeling biomedical texts at the same time as protein sequence modeling. 

%%%%%%%%%%%%%%%%%%%%%%%%%%%%%%%%%%%%%%%%%%%%%%%%%%%%%%%%%%%%

\subsection{Biomedical Language Models} \label{sec:pre:biomedical_lm}

Compared to the texts from general domains like newswire and Web, biomedical texts differ a lot in terms of vocabulary and expressions. 
To tackle such differences, language models specific to the biomedical domain~\cite{beltagy2019scibert,lee2020biobert,gu2021domain} are actively studied. 
% To tackle such differences, domain-specific language modeling strategies have been explored, including building biomedical text corpus~\cite{hunter2006biomedical}, biomedical vocabulary construction~\cite{gu2021domain} and training schedule design~\cite{lee2020biobert,gu2021domain}. 
In this work, we employ a performant biomedical language model, PubMedBERT~\cite{gu2021domain}, to represent the biomedical text descriptions of proteins. 
% We study how these text representations can benefit protein sequence pre-training and downstream protein understanding tasks. 

%%%%%%%%%%%%%%%%%%%%%%%%%%%%%%%%%%%%%%%%%%%%%%%%%%%%%%%%%%%%

%%%%%%%%%%%%%%%%%%%%%%%%%%%%%%%%%%%%%%%%%%%%%%%%%%%%%%%%%%%%

\section{Method} \label{sec:method}

In this section, we first motivate the proposed {\method} framework and present its general picture in Sec.~\ref{sec:method:overview}, and then elucidate the design of pre-training tasks in Sec.~\ref{sec:method:pretrain}, followed by discussing the connections with and advantages over previous works in Sec.~\ref{sec:method:discuss}. 

%%%%%%%%%%%%%%%%%%%%%%%%%%%%%%%%%%%%%%%%%%%%%%%%%%%%%%%%%%%%

\begin{figure*}[t]
\centering
    \includegraphics[width=1.0\linewidth]{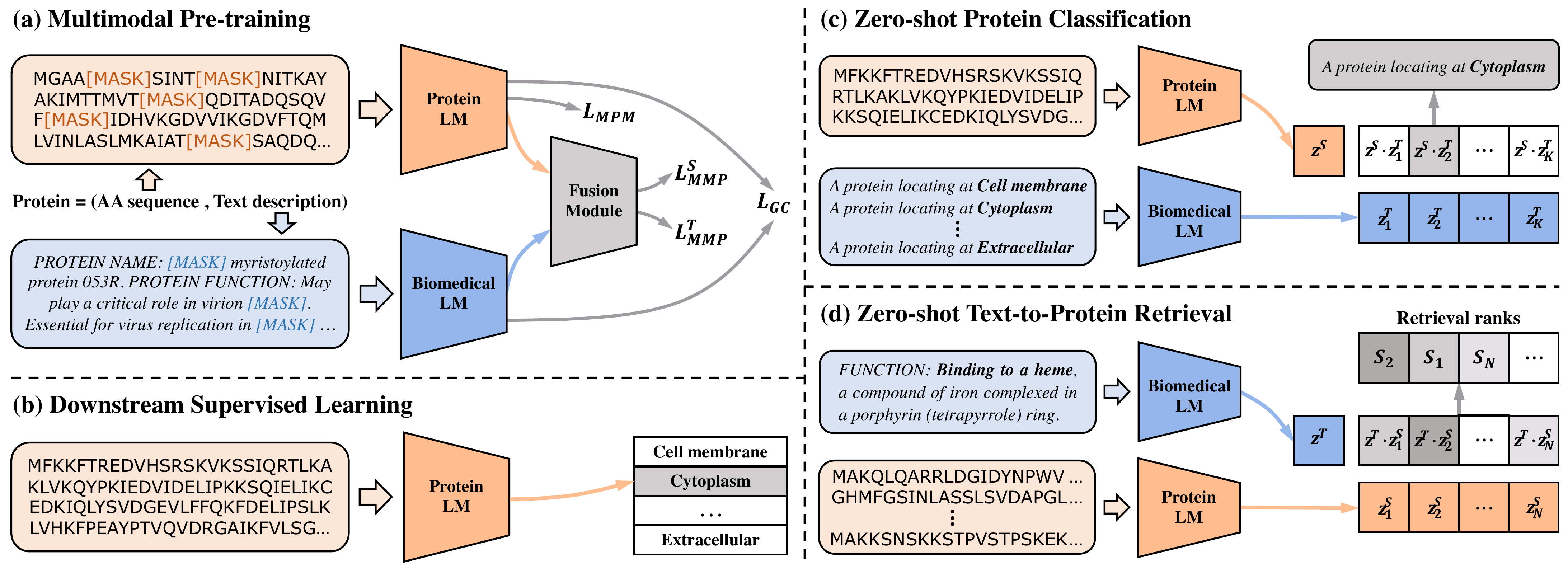}
    \vspace{-6.6mm}
    \caption{\textbf{Graphical illustration of {\method} framework}. (a) A protein language model (PLM) is first pre-trained along with a biomedical language model (BLM) and a fusion module to jointly model protein sequences and biomedical texts. (b)~After this multi-modal pre-training, the PLM can be used individually for supervised learning on downstream tasks. (c)~The couple of pre-trained PLM and BLM can perform zero-shot protein classification using only label descriptions. (d)~The paired PLM and BLM can also retrieve functional proteins from a large-scale database without any function annotation.}
    % \caption{\textbf{Graphical illustration of {\method} framework}. (a) During pre-training, a protein language model (PLM) is trained along with a biomedical language model (BLM) and a fusion module to jointly model protein sequences and biomedical language. (b) The pre-trained PLM can be used individually to perform supervised learning on downstream tasks. (c) The couple of pre-trained PLM and BLM can be used for zero-shot prediction based on their aligned representation space.}
    \label{fig:framework}
\end{figure*}

%%%%%%%%%%%%%%%%%%%%%%%%%%%%%%%%%%%%%%%%%%%%%%%%%%%%%%%%%%%%

\subsection{Motivation and Overview} \label{sec:method:overview}

\textbf{Motivation:} Existing PLMs~\citep{elnaggar2020prottrans,lu2020self,rives2021biological,lin2022language} learn protein representations primarily based on their sequences, which can well capture co-evolutionary information but cannot explicitly acquire various protein properties like protein functions and subcellular locations. By acquiring such property information, the effectiveness of a PLM can be further improved, considering that the protein properties studied in pre-training and downstream tasks can correlate with each other~\cite{bhardwaj2005correlation}.

To gain such improvement, we curate the {\dataset} dataset that augments protein sequences with text descriptions of their diverse properties (see Sec.~\ref{sec:exp:pretrain} for details). By injecting such property information into protein sequence representations, we aim at (1) a PLM that is more effective than previous ones on various downstream tasks under supervised learning, and (2) it can further enable zero-shot prediction through the generalization of text descriptions between known protein properties and unknown ones.

\textbf{{\method} Framework:} 
% To achieve this goal, we propose the {\method} framework with four main components (framework overview is shown in Fig.~\ref{fig:framework}): 
To attain these goals, we first perform multi-modal pre-training of sequences and texts and then apply the pre-trained model to three types of downstream applications (framework overview is shown in Fig.~\ref{fig:framework}):
% we perform one pre-training step followed by three types of downstream applications
% in which pre-training is performed by jointly modeling protein sequences and their corresponding text descriptions, and downstream tasks can be solved by supervised learning or zero-shot prediction. We present the graphical overview of this framework in Fig.~\ref{fig:framework} and clarify its four parts as below.
\vspace{-2.5mm}
\begin{itemize}
    \item \textbf{Multimodal Pre-training:} Given the {\dataset} dataset, we train a PLM together with a biomedical language model (BLM) and a fusion module to model the paired protein sequences and text descriptions. 
    We consider three kinds of pre-training tasks, \emph{i.e.}, unimodal mask prediction, multimodal representation alignment and multimodal mask prediction, to capture the protein property information with different granularities and also preserve the PLM's original representation power. 
    % We consider three kinds of pre-training tasks: (1) \emph{unimodal mask prediction}: this task recovers a corrupted protein sequence based only on itself, which models the data dependency within a single modality; (2) \emph{multimodal representation alignment}: this task seeks to align the representations of each paired protein sequence and text description, which forms a common latent space for two different modalities; (3) \emph{multimodal mask prediction}: this task recovers data corruption under the guidance of both protein sequence and text description, which models fine-grained cross-modality data dependency. 
    % We motivate these tasks and state the detailed design in Sec.~\ref{sec:method:pretrain}.  
    \vspace{-1.5mm}
    \item \textbf{Downstream Supervised Learning:} After such pre-training, the PLM is enriched by the useful property information within biomedical texts. For downstream tasks with labeled proteins, we can employ the PLM individually to solve the tasks by supervised learning.
    % scheme, where the PLM can be used as a feature extractor with pre-trained weights frozen or fine-tuned along with the task-specific prediction head. 
    % We study this type of downstream applications in Sec.~\ref{sec:exp:repr}. 
    \vspace{-1.5mm}
    \item \textbf{Zero-shot Protein Classification:} When a protein classification task occurs without any labeled data, {\method} enables zero-shot classification. Specifically, the classification result can be determined by the representation similarity comparison between the query protein and all labels, thanks to the aligned representation space of protein sequences and label descriptions. 
    % Specifically, we can embed the sequence of a query protein by the PLM and embed the text descriptions of all classification labels by the BLM, and the classification result is determined by the similarity comparison between the query protein representation and all label representations. 
    % Such a zero-shot predictor can also be used to enhance a supervised predictor (if some labels are available) by refining decision boundaries. 
    % The applications of such zero-shot classifiers are studied in Sec.~\ref{sec:exp:zero}.  
    \vspace{-1.5mm}
    \item \textbf{Zero-shot Text-to-Protein Retrieval:} Based on the aligned representation space, {\method} also allows us to retrieve functional proteins from a large-scale database by using only the text descriptions of protein functions, in which no function annotation is required. 
    % Such kind of applications are investigated in Sec.~\ref{sec:exp:t2p}. 
\end{itemize}

%%%%%%%%%%%%%%%%%%%%%%%%%%%%%%%%%%%%%%%%%%%%%%%%%%%%%%%%%%%%

\subsection{Pre-training Tasks: Joint Modeling of Protein Sequences and Biomedical Texts} \label{sec:method:pretrain}

During {\method} pre-training, we aim to learn informative protein sequence representations guided by biomedical texts. To start this process with decent representations of protein sequences and biomedical texts, we use pre-trained PLM (\emph{i.e.}, ProtBert~\cite{elnaggar2020prottrans}, ESM-1b~\cite{rives2021biological} or ESM-2~\cite{lin2022language}) and pre-trained BLM (\emph{i.e.}, PubMedBERT~\cite{gu2021domain}) for initialization. During training, we tune the parameters of PLM and freeze those of BLM, since the pre-trained BLM is sufficient for extracting semantically meaningful representations from biomedical texts, and it is computationally expensive to tune both PLM and BLM simultaneously. {\method} involves the following pre-training tasks for representation learning. 

\textbf{Unimodal Mask Prediction:} The PLM for initialization is pre-trained by masked protein modeling (MPM), \emph{i.e.}, predicting masked residues based on the protein sequence context. This task can capture co-evolutionary information by modeling residue type dependency. 
% and is demonstrated to be effective on boosting downstream protein understanding tasks~\cite{rao2019evaluating,xu2022peer}. 
To preserve such unimodal information when injecting the cross-modality information from biomedical texts, we keep an MPM loss function $\mathcal{L}_{\mathrm{MPM}}$ for {\method} pre-training. Specifically, for each protein sequence, we randomly mask 15\% residue tokens and predict each masked token based on its contextualized representation extracted by the PLM, where $\mathcal{L}_{\mathrm{MPM}}$ is formulated as a cross-entropy loss to measure the cost. 

\textbf{Multimodal Representation Alignment:} The biomedical text representations learned by a pre-trained BLM can well reflect the semantics of the texts~\cite{jin2019probing,gu2021domain}. 
% The pre-trained BLM can distribute biomedical text representations based on the semantics of the texts~\cite{jin2019probing,gu2021domain}. 
Therefore, when given protein property descriptions, the BLM can extract semantically meaningful text representations of proteins. 
% the BLM can distribute the proteins based on their property relevance. 
Thanks to this capability, by aligning protein sequence representations to their associated text representations, we can naturally inject protein property information into sequence representations. 

To realize such alignment, we perform contrastive learning between protein sequences and their text descriptions. Given a batch of $M$ proteins $\{P_i=(S_i, T_i)\}_{i=1}^M$, we use the PLM to extract protein sequence representations $\{z^S_i\}_{i=1}^M$ and the BLM to derive text description representations $\{z^T_i\}_{i=1}^M$. A standard InfoNCE loss~\cite{oord2018representation} $\mathcal{L}_{\mathrm{GC}}$ is defined to maximize the representation similarity between corresponding sequences and texts and minimize the similarity between negative pairs:
\begin{equation} \label{eq:infonce}
% \small
\begin{split}
\mathcal{L}_{\mathrm{GC}} = - \frac{1}{2M} \sum_{i=1}^M \Bigg ( & \log \frac{\exp(z^S_i \cdot z^T_i / \tau)}{\sum_{j=1}^M \exp(z^S_i \cdot z^T_j / \tau)} \\ 
& + \log \frac{\exp(z^S_i \cdot z^T_i / \tau)}{\sum_{j=1}^M \exp(z^S_j \cdot z^T_i / \tau)} \Bigg ) ,
\end{split}
\end{equation}
where, under multi-GPU data parallelism, 
we gather whole-batch samples separated on different GPUs to form negative pairs 
% we gather the samples from all GPUs to form negative pairs 
and thus term the loss $\mathcal{L}_{\mathrm{GC}}$ as a \emph{global contrastive (GC) loss} following the convention~\cite{singh2022flava}, and $\tau$ denotes a learnable temperature parameter. 

\textbf{Multimodal Mask Prediction:} Although the general dependency between the whole protein sequences and full text descriptions can be well modeled by $\mathcal{L}_{\mathrm{GC}}$, $\mathcal{L}_{\mathrm{GC}}$ alone does not capture the dependency between the residues in a protein sequence and the words in its text description. 
% how the residues in a protein sequence and the words in the protein's text description depend on each other. 
Such fine-grained cross-modality interdependency is actually ubiquitous. For example, \emph{a soluble protein} (descriptive words) always co-occurs with charged and polar surface residues~\cite{capaldi1972low}; 
\emph{high thermostability} (descriptive words) and high amounts of hydrophobic residues are correlated with each other~\cite{kumar2000factors}, \emph{etc.} 
% \emph{low protein expression} (descriptive words) interrelates with hydrophobic and aromatic residues~\cite{weber2020impact}, \emph{etc.} 
To capture such interdependency, we propose a novel pre-training task that encourages the model to recover the corrupted protein sequence (or text description) based on the information from both modalities. 

Specifically, given a protein sequence $S=[s_1, s_2, \cdots , s_n]$ and its corresponding text description $T=[t_1, t_2, \cdots , t_m]$, we first randomly mask 15\% residues in the protein sequence and 15\% words in the text description. Upon the corrupted inputs, we employ the PLM to extract residue representations $Z^S = [z^s_1, z^s_2, \cdots , z^s_n]$ and utilize the BLM to extract word representations $Z^T = [z^t_1, z^t_2, \cdots , z^t_m]$. A \textbf{fusion module} with both self- and cross-attention is then used to model the interdependency between residues and words, in which each residue and word updates its representation by attending to all the tokens along both protein sequence and text description (we state the detailed architecture in Appendix~\ref{supp:sec:pretrain_arch}). The fusion module produces the fused residue representations $\tilde{Z}^S = [\tilde{z}^s_1, \tilde{z}^s_2, \cdots , \tilde{z}^s_n]$ and the fused word representations $\tilde{Z}^T = [\tilde{z}^t_1, \tilde{z}^t_2, \cdots , \tilde{z}^t_m]$, in which each residue/word representation combines the information from both modalities. Based on $\tilde{Z}^S$ and $\tilde{Z}^T$, we perform \emph{multimodal mask prediction (MMP)} to recover masked residues and words, where a cross-entropy loss $\mathcal{L}_{\mathrm{MMP}}^S$ measures the cost on protein sequence, and another cross-entropy loss $\mathcal{L}_{\mathrm{MMP}}^T$ measures the cost on text description, inducing the overall MMP loss $\mathcal{L}_{\mathrm{MMP}} = \mathcal{L}_{\mathrm{MMP}}^S + \mathcal{L}_{\mathrm{MMP}}^T$. 

\textbf{Overall Pre-training Objective:} During the pre-training process, we seek to minimize the loss functions of all pre-training tasks simultaneously: 
\begin{equation} \label{eq:overall_obj}
    \min \limits_{\theta} \, \mathcal{L}_{\mathrm{MPM}} + \mathcal{L}_{\mathrm{GC}} + \mathcal{L}_{\mathrm{MMP}} ,
\end{equation}
where $\theta$ denotes all learnable parameters including those of the PLM, the fusion module and all projection/prediction heads. We state the detailed architectures of these modules in Appendix~\ref{supp:sec:pretrain_arch}. 

%%%%%%%%%%%%%%%%%%%%%%%%%%%%%%%%%%%%%%%%%%%%%%%%%%%%%%%%%%%%

\subsection{Discussion} \label{sec:method:discuss}

Now we discuss the connections of our method with previous works and emphasize its advantages.

\textbf{Advantages over Self-Supervised PLMs:} Previous self-supervised PLMs~\citep{elnaggar2020prottrans,rives2021biological,lin2022language} and the proposed {\method}-induced ones can both capture co-evolutionary information hidden in protein sequences by masked protein modeling. On this basis, {\method}-induced PLMs further utilize the supervision from textual protein property descriptions, and they are guided to acquire whole-protein properties by multimodal representation alignment and acquire residue-level properties by multimodal mask prediction. 

\textbf{Advantages over OntoProtein~\citep{zhang2022ontoprotein}:} Similar to our approach, OntoProtein also seeks to enhance a self-supervised PLM by involving protein property information. In comparison, {\method} could be more effective mainly in two aspects. (1) \textbf{Diversity of considered properties:} OntoProtein retrieves Gene Ontology terms~\citep{zhang2022ontoprotein} to cover protein functions and locations; besides these two kinds of properties, {\method} additionally includes protein names and families which are useful to indicate protein structural and functional similarity~\citep{murzin1995scop}. (2) \textbf{Property modeling manner:} OntoProtein learns a closed set of protein properties under the context of a \emph{fixed biological knowledge graph}, which limits its ability to generalize to unknown properties of new proteins, 
% in which the generalization towards unknown properties of new proteins is hard to be modeled, 
% OntoProtein applys a \emph{fixed biological knowledge graph} to represent a closed set of protein properties and thus can hardly generalize to unknown properties of new proteins, 
while {\method} can flexibly model such generalization based on the semantic correlation of text descriptions between known and unknown properties, leading to decent zero-shot prediction capability (studied in Secs.~\ref{sec:exp:zero} and \ref{sec:exp:t2p}).

\vspace{-0.8mm}
\section{Experiments} \label{sec:exp}
\vspace{-0.2mm}

%%%%%%%%%%%%%%%%%%%%%%%%%%%%%%%%%%%%%%%%%%%%%%%%%%%%%%%%%%%%

\begin{table}[t]
\begin{spacing}{1.05}
\centering
\caption{Statistics of the {\dataset} dataset.}
\vspace{-2.5mm}
\label{tab:protdescribe}
\begin{adjustbox}{max width=1\linewidth}
    \begin{tabular}{l|cccc}
        \toprule
        \bf{Field} & \bf{Name} & \bf{Function} & \bf{Location} & \bf{Family} \\
        \midrule
        \bf{\#Covered samples} & 553,052 & 460,936 & 350,929 & 512,276 \\
        \bf{Coverage} & 100\% & 83.3\% & 63.5\% & 92.6\% \\
        \bottomrule
    \end{tabular}
\end{adjustbox}
\end{spacing}
\vspace{-4mm}
\end{table}

%%%%%%%%%%%%%%%%%%%%%%%%%%%%%%%%%%%%%%%%%%%%%%%%%%%%%%%%%%%%

\subsection{Pre-training Setups} \label{sec:exp:pretrain}

\textbf{Pre-training Dataset:} To inject protein property information into PLMs, we build the {\dataset} dataset with 553,052 aligned pairs of protein sequence and property description. Specifically, we employ the Swiss-Prot~\cite{bairoch2000swiss} database to provide annotations of various protein properties, in which we select four property fields: (1) ``\emph{Protein Name}'' gives the full protein name recommended by the UniProt consortium~\cite{uniprot2019uniprot}; (2) ``\emph{Function}'' depicts diverse functions owned by a protein; (3) ``\emph{Subcellular Location}'' describes the location and topology of a mature protein in the cell; (4) ``\emph{Similarity}'' provides information about the protein families that a protein belongs to. A complete property description is formed by concatenating these four fields in order, where missing fields are skipped (see Appendix~\ref{supp:sec:setup:pretrain} for the detailed concatenation scheme and examples). Tab.~\ref{tab:protdescribe} presents the statistics of how each field covers the whole dataset. 

\textbf{Protein Language Models:} We seek to enhance three performant PLMs, \emph{i.e.}, ProtBert~\cite{elnaggar2020prottrans}, ESM-1b~\cite{rives2021biological} and ESM-2~\cite{lin2022language}, by tuning their weights through the proposed {\method} pre-training. We name the PLMs after this pre-training phase as \textbf{{\method}-ProtBert}, \textbf{{\method}-ESM-1b} and \textbf{{\method}-ESM-2}. For ProtBert, we employ the ProtBert-BFD version which is trained on the BFD database~\cite{steinegger2018clustering}. For ESM-2, we adopt the ESM-2-650M model so as to fairly compare with ESM-1b under the same model size. 

\textbf{Biomedical Language Models:} By default, we utilize the PubMedBERT-abs~\cite{gu2021domain} trained on PubMed abstracts to extract representations of protein property descriptions. We study another model version, PubMedBERT-full trained with additional full-text articles, in Appendix~\ref{supp:sec:ablation:blm}. 

% PubMedBERT has two model versions, \emph{i.e.}, (1) the PubMedBERT-abs which is trained by using only PubMed abstracts and (2) the PubMedBERT-full which adds PubMed Central full-text articles to the training set. By default, we use the PubMedBERT-abs for {\method} pre-training, and we study the effect of using these two model versions in Sec.~\ref{sec:exp:analysis:ablation}.  

\textbf{Training Configurations:} An Adam optimizer~\cite{kingma2014adam} (learning rate: $1.0 \times 10^{-5}$, weight decay: 0) is used to train the whole model for 20 epochs on 4 Tesla V100 GPUs. More settings are introduced in Appendix~\ref{supp:sec:setup:pretrain}. 

%%%%%%%%%%%%%%%%%%%%%%%%%%%%%%%%%%%%%%%%%%%%%%%%%%%%%%%%%%%%

\subsection{Representation Learning} \label{sec:exp:repr}

\subsubsection{Experimental Setups} \label{sec:exp:repr:setup}

\textbf{Downstream Benchmark Tasks.} We adopt 11 benchmark tasks within three task types (the ``\emph{Abbr.}'' below denotes the abbreviated task name in Tab.~\ref{tab:loc-fit} and \ref{tab:anno}):
\vspace{-2.5mm}
\begin{itemize}
    \item \textbf{Protein Localization Prediction} seeks to predict the subcellular locations of proteins. We consider two such problems from DeepLoc~\cite{almagro2017deeploc}, the subcellular localization prediction (\emph{Abbr.}, Sub) with 10 location categories and the binary localization prediction (\emph{Abbr.}, Bin) with 2 location categories. We follow the official dataset splits. 
    % We adopt two such problems from DeepLoc~\cite{almagro2017deeploc}, \emph{i.e.}, the \emph{subcellular localization prediction} task that classifies proteins into 10 location categories and the \emph{binary localization prediction} task that classifies proteins into ``membrane-bound'' or ``soluble''. We follow the official dataset splits for these two tasks. 
    \vspace{-1.5mm}
    \item \textbf{Fitness Landscape Prediction} aims to predict the effect of residue mutations on protein fitness. We employ the $\beta$-lactamase (\emph{Abbr.}, $\beta$-lac) landscape from PEER~\cite{xu2022peer}, the AAV and Thermostability (\emph{Abbr.}, Thermo) landscapes from FLIP~\cite{dallago2021flip}, and the Fluorescence (\emph{Abbr.}, Flu) and Stability (\emph{Abbr.}, Sta) landscapes from TAPE~\cite{rao2019evaluating}. For AAV, we use the ``two\_vs\_many'' dataset splits; for Thermostability, we adopt the ``human\_cell'' splits; we follow the only default splits on all other tasks. In Appendix~\ref{supp:sec:proteingym}, we further show the results on ProteinGym~\cite{notin2022tranception}. 
    % In our experiments, we employ the $\beta$-lactamase landscape from PEER~\cite{xu2022peer}, the AAV and Thermostability landscapes from FLIP~\cite{dallago2021flip}, and the Fluorescence and Stability landscapes from TAPE~\cite{rao2019evaluating}. For AAV, we use the ``two\_vs\_many'' dataset splits; for Thermostability, we adopt the ``human\_cell'' dataset splits; we follow the official dataset splits for all other tasks. 
    \vspace{-1.5mm}
    \item \textbf{Protein Function Annotation} seeks to annotate a protein with multiple functional labels. We employ two standard benchmarks proposed by DeepFRI~\cite{gligorijevic2021structure}, \emph{i.e.}, Enzyme Commission (EC) number prediction and Gene Ontology (GO) term prediction. The GO benchmark is split into three branches to predict molecular function (\emph{Abbr.}, GO-MF), biological process (\emph{Abbr.}, GO-BP) and cellular component (\emph{Abbr.}, GO-CC). Following \citet{zhang2022protein}, we use the dataset splits under 95\% sequence identity cutoff for both EC and GO. 
    % The EC prediction task annotates enzymes according to the biochemical reactions catalyzed by them. The GO prediction task is split into three branches based on three ontologies: molecular function (MF), biological process (BP) and cellular component (CC). Following GearNet~\cite{zhang2022protein}, we use the dataset splits under 95\% sequence identity cutoff for both benchmarks.  
\end{itemize}

\textbf{Baselines:} We adopt four protein sequence encoders trained from scratch, \emph{i.e.}, CNN~\cite{shanehsazzadeh2020transfer}, ResNet~\cite{rao2019evaluating}, LSTM~\cite{rao2019evaluating} and Transformer~\cite{rao2019evaluating}, as naive baselines. We focus on comparing with four performant PLMs, \emph{i.e.}, ProtBert~\cite{elnaggar2020prottrans}, OntoProtein~\cite{zhang2022ontoprotein}, ESM-1b~\cite{rives2021biological} and ESM-2~\cite{lin2022language}.

\textbf{Training and Evaluation:} We train with an Adam optimizer for 100 epochs on localization and fitness prediction tasks and for 50 epochs on function annotation tasks. For localization and fitness prediction, all PLMs are evaluated under both fix-encoder learning and full-model tuning settings, and only full-model tuning is used for PLMs on function annotation, since it is hard to solve the multiple binary classification problems on EC and GO with fixed protein representations. More training details are stated in Appendix~\ref{supp:sec:setup:repr}.

For all models on all tasks, we select the checkpoint for evaluation based on the validation set performance, and all results are reported on the seed 0. We measure the classification accuracy for localization prediction and the Spearman's $\rho$ for fitness prediction. Following \citet{gligorijevic2021structure}, function annotation tasks are measured by AUPR and $\mathrm{F}_{\mathrm{max}}$ whose detailed definitions are in Appendix~\ref{supp:sec:setup:repr}. 
% We provide the detailed definitions of AUPR and $\mathrm{F}_{\mathrm{max}}$ in Appendix~\ref{supp:sec:setup:repr}. 

%%%%%%%%%%%%%%%%%%%%%%%%%%%%%%%%%%%%%%%%%%%%%%%%%%%%%%%%%%%%

\begin{table}[t]
\begin{spacing}{1.05}
\centering
\caption{Benchmark results on protein localization and fitness landscape prediction. We use three color scales of blue to denote the \textcolor{r1}{first}, \textcolor{r2}{second} and \textcolor{r3}{third} best performance. \emph{Abbr.}, Loc.: Localization; pred.: prediction; Acc: accuracy.}
\vspace{-2.5mm}
\label{tab:loc-fit}
\begin{adjustbox}{max width=1\linewidth}
% \begin{threeparttable}
    \begin{tabular}{l|cc|ccccc|c}
        \toprule
        \multirow{2}{*}{\bf{Model}} & \multicolumn{2}{c|}{\bf{Loc. pred.} (\emph{Acc\%})} &
        \multicolumn{6}{c}{\bf{Fitness pred.} (\emph{Spearman's \bm{$\rho$}})} \\
        \cmidrule{2-3}
        \cmidrule{4-9}
        & \bf{Bin} & \bf{Sub} & \bf{$\beta$-lac} & \bf{AAV} & \bf{Thermo} & \bf{Flu} & \bf{Sta} & \bf{Mean \bm{$\rho$}} \\
        \midrule
        \multicolumn{9}{c}{\bf{Protein sequence encoders trained from scratch}} \\
        \midrule
        CNN & 82.67 & 58.73 & 0.781 & 0.746 & 0.494 & \cellcolor{r1} \textbf{0.682} & 0.637 & 0.668 \\
        ResNet & 78.99 & 52.30 & 0.152 & 0.739 & 0.528 & 0.636 & 0.126 & 0.436 \\
        LSTM & 88.11 & 62.98 & 0.139 & 0.125 & 0.564 & 0.494 & 0.533 & 0.371 \\
        Transformer & 75.74 & 56.02 & 0.261 & 0.681 & 0.545 & 0.643 & 0.649 & 0.556 \\
        \midrule
        \multicolumn{9}{c}{\bf{PLMs \emph{w/} fix-encoder learning}} \\
        \midrule
        ProtBert & 81.54 & 59.44 & 0.616 & 0.209 & 0.562 & 0.339 & 0.697 & 0.485 \\
        OntoProtein & 84.87 & 68.34 & 0.471 & 0.217 & 0.605 & 0.432 & 0.688 & 0.483 \\
        ESM-1b & 91.61 & 79.82 & 0.528 & 0.454 & 0.674 & 0.430 & 0.750 & 0.567 \\
        ESM-2 & 91.32 & \cellcolor{r3} 80.84 & 0.559 & 0.374 & 0.677 & 0.456 & 0.746 & 0.562 \\
        \bf{{\method}-ProtBert} & 92.29 & 78.49 & 0.569 & 0.219 & 0.621 & 0.376 & 0.719 & 0.501 \\
        \bf{{\method}-ESM-1b} & \cellcolor{r1} \textbf{92.87} & \cellcolor{r2} 82.00 & 0.578 & 0.460 & \cellcolor{r3} 0.680 & 0.523 & \cellcolor{r3} 0.766 & 0.601 \\
        \bf{{\method}-ESM-2} & \cellcolor{r2} 92.52 & \cellcolor{r1} \textbf{83.39} & 0.565 & 0.398 & \cellcolor{r2} 0.681 & 0.499 & \cellcolor{r1} \textbf{0.776} & 0.584 \\
        \midrule
        \multicolumn{9}{c}{\bf{PLMs \emph{w/} full-model tuning}} \\
        \midrule
        ProtBert & 91.32 & 76.53 & 0.731 & 0.794 & 0.660 & \cellcolor{r2} 0.679 & \cellcolor{r2} 0.771 & 0.727 \\
        OntoProtein & \cellcolor{r3} 92.47 & 77.59 & 0.757 & 0.791 & 0.662 & 0.630 & 0.731 & 0.714 \\
        ESM-1b & 92.40 & 78.13 & 0.839 & \cellcolor{r3} 0.821 & 0.669 & \cellcolor{r2} 0.679 & 0.694 & 0.740 \\
        ESM-2 & 91.72 & 78.67 & \cellcolor{r3} 0.867 & 0.817 & 0.672 & \cellcolor{r3} 0.677 & 0.718 & 0.750 \\
        \bf{{\method}-ProtBert} & 91.78 & 78.71 & 0.863 & 0.804 & 0.673 & \cellcolor{r2} 0.679 & 0.745 & \cellcolor{r3} 0.753 \\
        \bf{{\method}-ESM-1b} & 92.35 & 78.73 & \cellcolor{r1} \textbf{0.895} & \cellcolor{r1} \textbf{0.850} & \cellcolor{r2} 0.681 & \cellcolor{r1} \textbf{0.682} & 0.751 & \cellcolor{r1} \textbf{0.772} \\
        \bf{{\method}-ESM-2} & \cellcolor{r2} 92.52 & 80.22 & \cellcolor{r2} 0.879 & \cellcolor{r2} 0.825 & \cellcolor{r1} \textbf{0.682} & \cellcolor{r1} \textbf{0.682} & 0.738 & \cellcolor{r2} 0.761 \\
        \bottomrule
    \end{tabular}
% \begin{tablenotes}
%     \item[*] Fix-encoder learning: the PLM is used as a feature extractor with pre-trained weights frozen.
% \end{tablenotes}
% \end{threeparttable}
\end{adjustbox}
\end{spacing}
\vspace{-5mm}
\end{table}

%%%%%%%%%%%%%%%%%%%%%%%%%%%%%%%%%%%%%%%%%%%%%%%%%%%%%%%%%%%%

\subsubsection{Experimental Results} \label{sec:exp:repr:result} 

We report the benchmark results on localization and fitness prediction in Tab.~\ref{tab:loc-fit} and report function annotation results in Tab.~\ref{tab:anno}. Based on the benchmark results, we have the following observations:

\textbf{{\method}-induced PLMs clearly outperform the vanilla PLMs.} It is observed that: (1) {\method}-ProtBert outperforms the vanilla ProtBert on 21 out of 24 benchmark metrics (including both fix-encoder learning and full-model tuning ones); (2) {\method}-ESM-1b surpasses the vanilla ESM-1b on 22 out of 24 benchmark metrics; (3) {\method}-ESM-2 outperforms the vanilla ESM-2 on all 24 benchmark metrics. These results demonstrate that {\method} pre-training is generally beneficial to different PLMs, which boosts their performance on diverse downstream tasks. 

\textbf{{\method}-ProtBert performs consistently better than OntoProtein under fair comparison.} {\method}-ProtBert and OntoProtein can be fairly compared with each other, since they both adopt ProtBert as the initial PLM. {\method}-ProtBert surpasses OntoProtein on 22 out of 24 benchmark metrics, which verifies the superiority of the proposed pre-training dataset and pre-training tasks.  

\textbf{{\method}-ESM-1b performs best on fitness prediction, and {\method}-ESM-2 performs best on localization prediction and function annotation.} We can observe that: (1) {\method}-ESM-1b achieves the best performance on 4 out of 6 benchmark metrics for fitness prediction; (2) {\method}-ESM-2 obtains the highest localization prediction accuracy on average, and it performs best on 7 out of 8 benchmark metrics for function annotation. We therefore recommend these two PLMs as new state-of-the-arts. 

%%%%%%%%%%%%%%%%%%%%%%%%%%%%%%%%%%%%%%%%%%%%%%%%%%%%%%%%%%%%

\begin{table}[t]
\begin{spacing}{1.07}
\centering
\caption{Benchmark results on protein function annotation. We use three color scales of blue to denote the \textcolor{r1}{first}, \textcolor{r2}{second} and \textcolor{r3}{third} best performance.}
\vspace{-2.5mm}
\label{tab:anno}
\begin{adjustbox}{max width=1\linewidth}
    \begin{tabular}{l|cc|cc|cc|cc}
        \toprule
        \multirow{2}{*}{\bf{Model}} & \multicolumn{2}{c|}{\bf{EC}} & \multicolumn{2}{c|}{\bf{GO-BP}} & \multicolumn{2}{c|}{\bf{GO-MF}} & \multicolumn{2}{c}{\bf{GO-CC}} \\
        \cmidrule{2-3}
        \cmidrule{4-5}
        \cmidrule{6-7}
        \cmidrule{8-9}
        & AUPR & $\mathrm{F}_{\mathrm{max}}$ & AUPR & $\mathrm{F}_{\mathrm{max}}$ & AUPR & $\mathrm{F}_{\mathrm{max}}$ & AUPR & $\mathrm{F}_{\mathrm{max}}$ \\
        \midrule
        \multicolumn{9}{c}{\bf{Protein sequence encoders trained from scratch}} \\
        \midrule
        CNN & 0.540 & 0.545 & 0.165 & 0.244 & 0.380 & 0.354 & 0.261 & 0.387 \\
        ResNet & 0.137 & 0.187 & 0.166 & 0.280 & 0.281 & 0.267 & 0.266 & 0.403 \\
        LSTM & 0.032 & 0.082 & 0.130 & 0.248 & 0.100 & 0.166 & 0.150 & 0.320 \\
        Transformer & 0.187 & 0.219 & 0.135 & 0.257 & 0.172 & 0.240 & 0.170 & 0.380 \\
        \midrule
        \multicolumn{9}{c}{\bf{PLMs \emph{w/} full-model tuning}} \\
        \midrule
        ProtBert & 0.859 & 0.838 & 0.188 & 0.279 & 0.464 & 0.456 & 0.234 & 0.408 \\
        OntoProtein & 0.854 & 0.841 & 0.284 & 0.436 & 0.603 & 0.631 & 0.300 & 0.441 \\
        ESM-1b & 0.884 & \cellcolor{r3} 0.869 & \cellcolor{r3} 0.332 & 0.452 & 0.630 & 0.659 & \cellcolor{r3} 0.324 & \cellcolor{r3} 0.477 \\
        ESM-2 & \cellcolor{r3} 0.888 & \cellcolor{r2} 0.874 & \cellcolor{r2} 0.340 & \cellcolor{r3} 0.472 & \cellcolor{r3} 0.643 & \cellcolor{r2} 0.662 & \cellcolor{r2} 0.350 & 0.472 \\
        \bf{{\method}-ProtBert} & 0.876 & 0.856 & 0.286 & 0.440 & 0.615 & 0.648 & 0.314 & 0.449 \\
        \bf{{\method}-ESM-1b} & \cellcolor{r2} 0.894 & \cellcolor{r1} \textbf{0.878} & 0.328 & \cellcolor{r2} 0.480 & \cellcolor{r2} 0.644 & \cellcolor{r3} 0.661 & \cellcolor{r1} \textbf{0.364} & \cellcolor{r1} \textbf{0.488} \\
        \bf{{\method}-ESM-2} & \cellcolor{r1} \textbf{0.898} & \cellcolor{r1} \textbf{0.878} & \cellcolor{r1} \textbf{0.342} & \cellcolor{r1} \textbf{0.482} & \cellcolor{r1} \textbf{0.647} & \cellcolor{r1} \textbf{0.668} & \cellcolor{r1} \textbf{0.364} & \cellcolor{r2} 0.487 \\
        \bottomrule
    \end{tabular}
\end{adjustbox}
\end{spacing}
\vspace{-2mm}
\end{table}

%%%%%%%%%%%%%%%%%%%%%%%%%%%%%%%%%%%%%%%%%%%%%%%%%%%%%%%%%%%%

\subsection{Zero-shot Protein Classification} \label{sec:exp:zero}

\subsubsection{Experimental Setups} \label{sec:exp:zero:setup}

\textbf{Zero-shot Protein Classification based on Aligned Representation Space:} A {\method}-induced PLM naturally allows zero-shot protein classification, thanks to its aligned representation space of protein sequences and text descriptions. In specific, given the sequence $S$ of a query protein and the label descriptions $\{T_i\}_{i=1}^K$ of all $K$ classes, we employ the PLM to extract protein representation $z^S$ and use the jointly learned BLM to extract label representations $\{z^T_i\}_{i=1}^K$. We then derive classification logits $\{y_i\}_{i=1}^K$ by comparing the dot product similarity between protein and label representations: $y_i = z^S \cdot z^T_i / \tau$ ($i=1, \cdots, K$), which follows the formula of InfoNCE loss in Eq.~(\ref{eq:infonce}). Softmax is performed upon these logits to derive classification probabilities. 

\textbf{Benchmark Tasks:} In this part of experiments, we adopt two protein classification tasks as benchmarks: (1) the \emph{subcellular localization prediction} task which is same as the one introduced in Sec.~\ref{sec:exp:repr:setup}; (2) the \emph{reaction classification} task proposed by \citet{hermosilla2020intrinsic} which reformulates the EC number prediction task introduced in Sec.~\ref{sec:exp:repr:setup} as a classification task with 384 reaction classes. We follow the official dataset splits for both tasks. 

\textbf{Prompt Engineering:} To extract discriminative label representations, we have tried three types of prompt templates to describe protein function/location labels. (1) \emph{Name only}: a label is described only by the name of a function or location (\emph{e.g.}, ``Cytoplasm''); (2) \emph{Natural language}: the name is embedded into a natural language template (\emph{e.g.}, ``A protein locating at Cytoplasm''); (3) \emph{Pre-training template}: the name is embedded into the template used during {\method} pre-training (\emph{e.g.}, ``SUBCELLULAR LOCATION: Cytoplasm''). The pre-training template is empirically verified to be more effective than other two templates, and thus it is used across all experiments of this section. The comparisons among these templates are provided in Appendix~\ref{supp:sec:setup:zero}. 

%%%%%%%%%%%%%%%%%%%%%%%%%%%%%%%%%%%%%%%%%%%%%%%%%%%%%%%%%%%%

\begin{figure}[t]
\centering
    \includegraphics[width=1.0\linewidth]{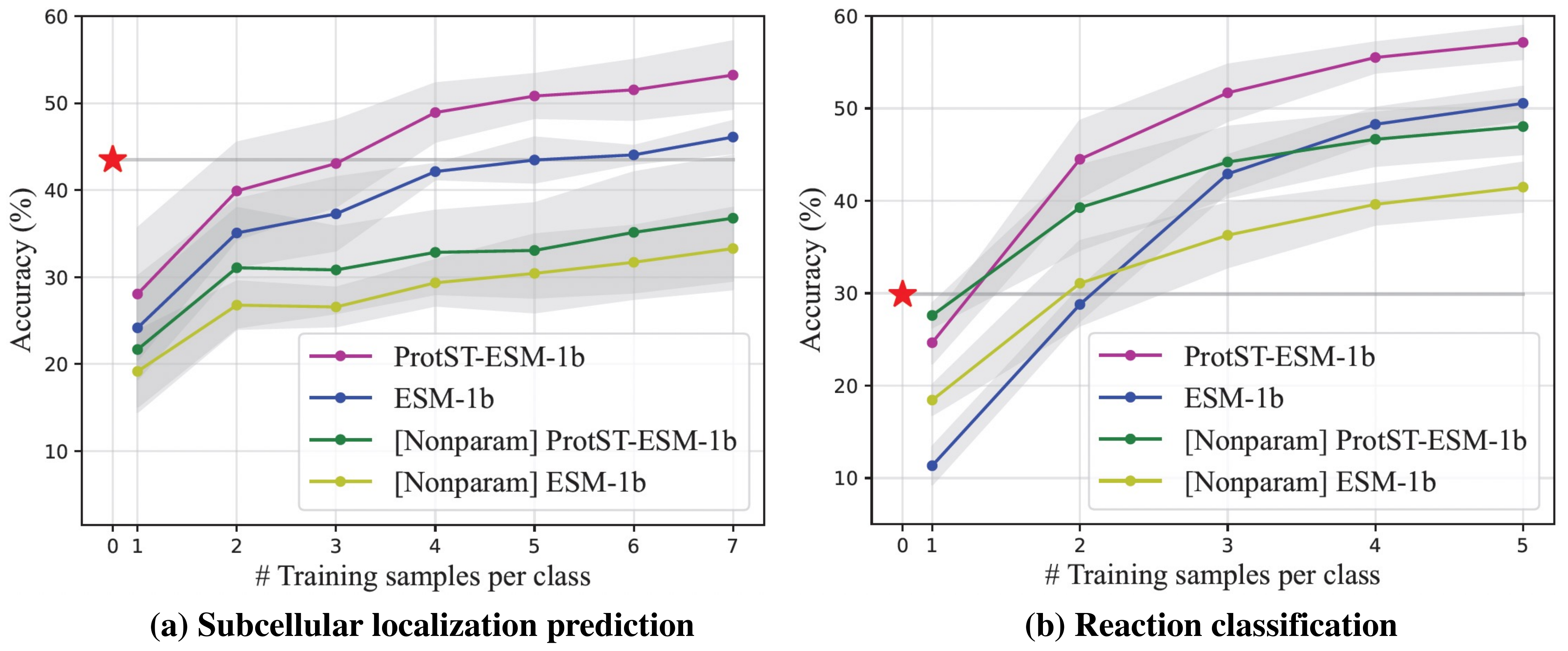}
    \vspace{-6.9mm}
    \caption{\textbf{Zero-shot {\method}-ESM-1b outperforms few-shot classifiers.} The horizontal line with a \textcolor{red}{red star} denotes the zero-shot performance of {\method}-ESM-1b. All few-shot results are averaged over seeds 0, 1, 2, 3 and 4, and gray intervals denote standard deviations.}
    \label{fig:data_efficiency}
    \vspace{-2.5mm}
\end{figure}

%%%%%%%%%%%%%%%%%%%%%%%%%%%%%%%%%%%%%%%%%%%%%%%%%%%%%%%%%%%%

\vspace{-0.2mm}
\subsubsection{Data Efficiency of Zero-shot Classifier} \label{sec:exp:zero:eff}
\vspace{-0.2mm}

\textbf{Baselines:} We study the data efficiency of zero-shot {\method}-ESM-1b by comparing it with $n$-shot classifiers ($n \geqslant 1$) which employ $n$ training samples per class for prediction. 
% Based on these training samples, 
We adopt four baselines: (1) the {\method}-ESM-1b with supervised fine-tuning, (2) the ESM-1b with supervised fine-tuning, (3) the nonparametric {\method}-ESM-1b classifier, and (4) the nonparametric ESM-1b classifier. We follow \citet{khandelwal2019generalization} to design the nonparametric classifiers 
% that well fit the few-shot setting, detailed in Appendix~\ref{supp:sec:setup:zero}. 
which predict based on the relations between test sample and training samples, and they well fit the few-shot prediction setting. 
We elucidate such classifiers in Appendix~\ref{supp:sec:setup:zero}. 
% All few-shot results are averaged over seeds 0, 1, 2, 3 and 4, and standard deviations are denoted by gray intervals in Fig.~\ref{fig:data_efficiency}. 

\textbf{Results:} For subcellular localization prediction (Fig.~\ref{fig:data_efficiency}(a)), the zero-shot {\method}-ESM-1b matches the performance of 3-shot supervised {\method}-ESM-1b and the performance of 5-shot supervised ESM-1b, and the zero-shot classifier outperforms two 7-shot nonparametric classifiers. For reaction classification (Fig.~\ref{fig:data_efficiency}(b)), the zero-shot {\method}-ESM-1b surpasses the 1-shot performance of supervised and nonparametric {\method}-ESM-1b, and it aligns the 2-shot performance of supervised and nonparametric ESM-1b. These results demonstrate the data efficiency of {\method}-induced zero-shot classifiers. In particular, they can be helpful in the downstream tasks with limited or even no labeled proteins by making educated predictions using only label descriptions. 

%%%%%%%%%%%%%%%%%%%%%%%%%%%%%%%%%%%%%%%%%%%%%%%%%%%%%%%%%%%%

\vspace{-0.5mm}
\subsubsection{Enhancing Supervised Learning with Zero-shot Classifier} \label{sec:exp:zero:enhance}

\textbf{Ensemble of Supervised Learning Model and Zero-shot Classifier:} We study how zero-shot {\method}-ESM-1b can boost supervised learning models via ensemble. 
Specifically, we combine the classification logits produced by a supervised learning model and the zero-shot classification logits as below: $\{y_k = y^{\mathrm{sup}}_k + \alpha\;\! y^{\mathrm{zero}}_k\}_{k=1}^K$ ($K$ is the number of classes), where $\alpha$ controls the contribution of the zero-shot classifier. 
% Given a test protein $S$, we use both a model trained with label supervision and the zero-shot classifier to compute its classification logits before softmax, denoted as $\{y^{\mathrm{sup}}_k\}_{k=1}^K$ and $\{y^{\mathrm{zero}}_k\}_{k=1}^K$ ($K$ is the number of classes). Ensemble is then performed upon these logits to get the final prediction: $\{y_k = y^{\mathrm{sup}}_k + \alpha\;\! y^{\mathrm{zero}}_k\}_{k=1}^K$, where $\alpha$ controls the contribution of the zero-shot classifier. 
Empirically, we set $\alpha$ as the ratio of the zero-shot classifier's validation set performance over the validation performance of the supervised learning model. 

\textbf{Baselines:} We employ {\method}-ESM-1b and ESM-1b with supervised fine-tuning on downstream tasks as baselines. We consider fine-tuning under both the few-shot setting and the full-shot setting (\emph{i.e.}, trained with all training samples). 
Based on these supervised models, we seek to utilize zero-shot {\method}-ESM-1b to enhance their performance. 

\begin{figure}[t]
\centering
    \includegraphics[width=1.0\linewidth]{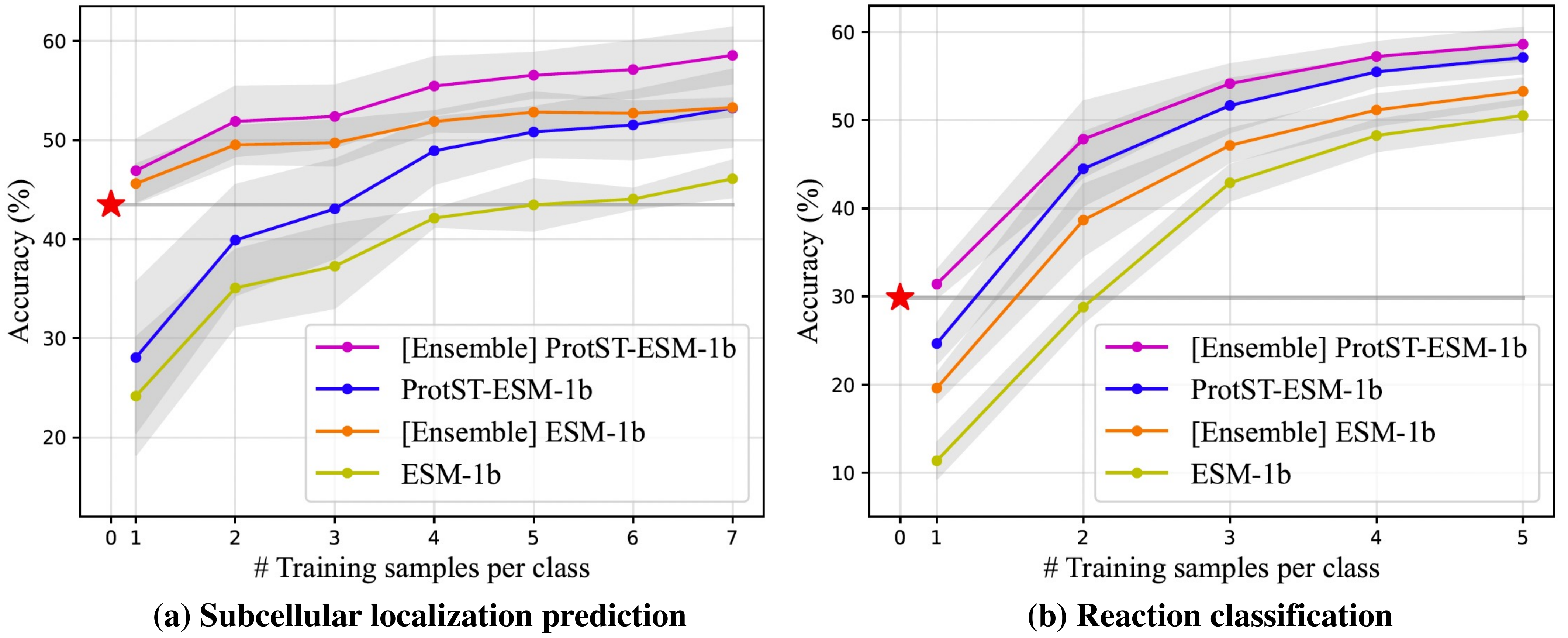}
    \vspace{-6.9mm}
    \caption{\textbf{Zero-shot {\method}-ESM-1b enhances few-shot classifiers' performance via ensemble.} The horizontal line with a \textcolor{red}{red star} denotes the zero-shot performance of {\method}-ESM-1b. All few-shot results are averaged over seeds 0, 1, 2, 3 and 4, and gray intervals denote standard deviations.}
    \label{fig:enhance_few_shot}
\vspace{-3.1mm}
\end{figure}

\textbf{Results:} According to Fig.~\ref{fig:enhance_few_shot} and Tab.~\ref{tab:enhance_full_shot}, we can observe that zero-shot {\method}-ESM-1b succeeds in enhancing the performance of all few-shot and full-shot baselines on both benchmarks. These results verify that {\method}-induced zero-shot classifiers are useful tools to enhance supervised learning models, which is realized by refining decision boundaries. 

%%%%%%%%%%%%%%%%%%%%%%%%%%%%%%%%%%%%%%%%%%%%%%%%%%%%%%%%%%%%

%%%%%%%%%%%%%%%%%%%%%%%%%%%%%%%%%%%%%%%%%%%%%%%%%%%%%%%%%%%%

\begin{table}[t]
\begin{spacing}{1.0}
\centering
\caption{\textbf{Zero-shot {\method}-ESM-1b enhances full-shot classifiers' performance via ensemble.} \emph{Abbr.}, loc.: localization; Acc: accuracy.}
\vspace{-2.5mm}
\label{tab:enhance_full_shot}
\begin{adjustbox}{max width=1.0\linewidth}
    \begin{tabular}{l|cc}
        \toprule
        \bf{Model} & \bf{Subcellular loc. (\emph{Acc\%})} & \bf{Reaction (\emph{Acc\%})} \\
        \midrule
        {\method}-ESM-1b & 82.00 & 86.73 \\
        \bf{[Ensemble] {\method}-ESM-1b} & \bf{82.37} & \bf{87.14} \\
        \midrule
        ESM-1b & 79.82 & 80.54 \\
        \bf{[Ensemble] ESM-1b} & \bf{80.20} & \bf{83.03} \\
        \bottomrule
    \end{tabular}
\end{adjustbox}
\end{spacing}
\vspace{-2.5mm}
\end{table}

%%%%%%%%%%%%%%%%%%%%%%%%%%%%%%%%%%%%%%%%%%%%%%%%%%%%%%%%%%%%

\begin{figure*}[t]
\centering
    \includegraphics[width=0.96\linewidth]{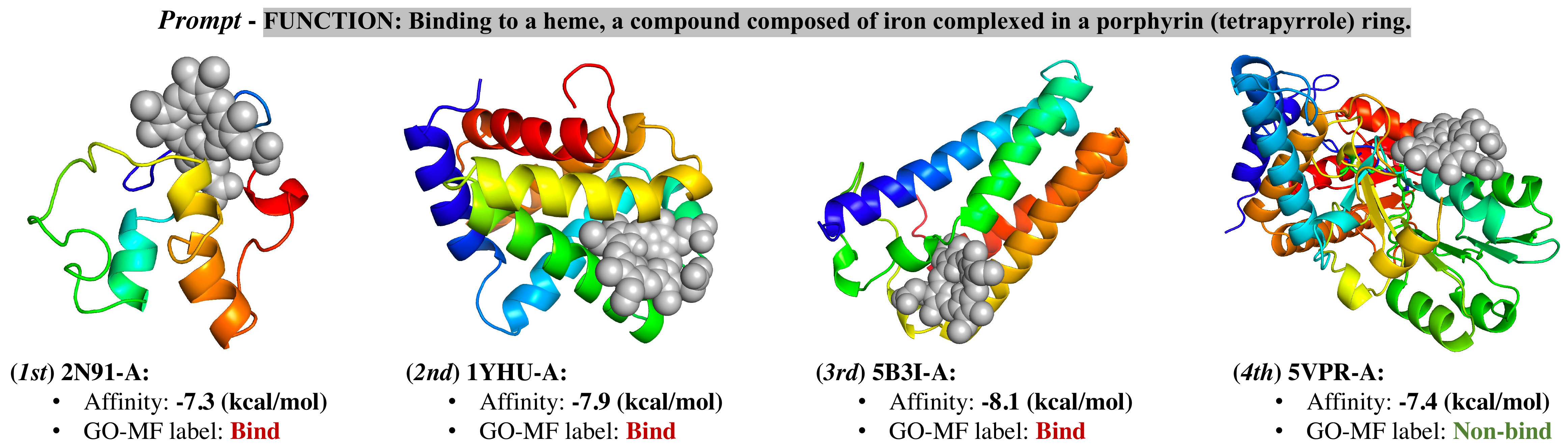}
    \vspace{-2.8mm}
    \caption{Zero-shot text-to-protein retrieval of heme binders based on {\method}-ESM-1b.}
    \label{fig:t2p}
\vspace{-2.5mm}
\end{figure*}

%%%%%%%%%%%%%%%%%%%%%%%%%%%%%%%%%%%%%%%%%%%%%%%%%%%%%%%%%%%%

\subsection{Zero-shot Text-to-Protein Retrieval} \label{sec:exp:t2p}

\textbf{Zero-shot Text-to-Protein Retriever:} Based on the protein-text aligned representation space, {\method} enables us to retrieve functional proteins from a large-scale database without any function annotation. To be specific, the PLM is first employed to extract the representations $\{z^S_i\}_{i=1}^N$ of all proteins in the database. During the retrieval process, given the text description (\emph{i.e.}, prompt) $T$ of a protein function, the BLM is used to extract its representation $z^T$, and all proteins are then ranked based on their representation similarity $\{\epsilon_i = z^S_i \cdot z^T\}_{i=1}^N$ with the prompt. 

\textbf{Experimental Setups:} We use {\method}-ESM-1b to retrieve the Gene Ontology (GO) dataset introduced in Sec.~\ref{sec:exp:repr:setup}. We build each prompt by adding the ``FUNCTION:'' prefix before the molecular function definition from GO. 

\textbf{Results:} In Fig.~\ref{fig:t2p}, we visualize the top-4 retrieved candidates of heme binders. We present the text prompt, the docking result of each candidate binding with heme (AutoDock Vina~\cite{trott2010autodock} is used for docking), the binding affinity predicted by AutoDock Vina (the lower the better), and the GO molecular function labels of heme binding. We can observe that the top-3 candidates are annotated as heme binders by GO, and the 4th candidate owns decent binding affinity though annotated as non-binding (only 0.54\% proteins are annotated as heme binders in the GO dataset). These results verify the effectiveness of {\method}-ESM-1b on retrieving heme binders. We provide more case studies in Appendix~\ref{supp:sec:t2p}. Other visualization results are in Appendix~\ref{supp:sec:visualize}. 

%%%%%%%%%%%%%%%%%%%%%%%%%%%%%%%%%%%%%%%%%%%%%%%%%%%%%%%%%%%%

\begin{table}[t]
\begin{spacing}{1.05}
\centering
\caption{\small Swiss-Prot \emph{v.s.} TrEMBL on protein property coverage.}
\vspace{-2.5mm}
\label{tab:compare_coverage}
\begin{adjustbox}{max width=0.85\linewidth}
    \begin{tabular}{l|cccc}
        \toprule
        \bf{Dataset} & \bf{Name} & \bf{Function} & \bf{Location} & \bf{Family} \\
        \midrule
        \bf{Swiss-Prot} & 100\% & 83.3\% & 63.5\% & 92.6\% \\
        \bf{TrEMBL} & 100\% & 24.0\% & 51.5\% & 78.0\% \\
        \bottomrule
    \end{tabular}
\end{adjustbox}
\end{spacing}
\vspace{-4mm}
\end{table}

%%%%%%%%%%%%%%%%%%%%%%%%%%%%%%%%%%%%%%%%%%%%%%%%%%%%%%%%%%%%

\begin{table}[t]
\begin{spacing}{1.1}
\centering
\caption{\small Swiss-Prot \emph{v.s.} TrEMBL as pre-training data source, compared on downstream representation learning tasks. \emph{Abbr.}, Loc.: localization prediction; Fit.: fitness prediction; Fix-enc.: fix-encoder learning; Full-m.: full-model tuning.}
\vspace{-1.5mm}
\label{tab:compare_pretrain_source}
\begin{adjustbox}{max width=0.85\linewidth}
    \begin{tabular}{l|cc|cc}
        \toprule
        \multirow{2}{*}{\bf{Dataset}} & \multicolumn{2}{c|}{\bf{Loc. (mean Acc\%)}} & \multicolumn{2}{c}{\bf{Fit. (mean \bm{$\rho$}})} \\
        \cmidrule{2-3}
        \cmidrule{4-5}
        & Fix-enc. & Full-m. & Fix-enc. & Full-m. \\
        \midrule
        \bf{Swiss-Prot} & \bf{87.44} & \bf{85.54} & \bf{0.601} & \bf{0.772} \\
        \bf{TrEMBL} & 86.68 & 85.13 & 0.597 & 0.762 \\
        \bottomrule
    \end{tabular}
\end{adjustbox}
\end{spacing}
\vspace{-2.5mm}
\end{table}

%%%%%%%%%%%%%%%%%%%%%%%%%%%%%%%%%%%%%%%%%%%%%%%%%%%%%%%%%%%%

\begin{table}[t]
\begin{spacing}{1.1}
\centering
\caption{\small Ablation study of pre-training losses on {\method}-ESM-1b. \emph{Abbr.}, Loc.: localization prediction; Fit.: fitness prediction; Func.: function annotation; Fix-enc.: fix-encoder learning; Full-m.: full-model tuning. \textcolor{r1}{Blue} denotes the largest decay.}
\vspace{-2.5mm}
\label{tab:ablation}
\begin{adjustbox}{max width=1.0\linewidth}
    \begin{tabular}{l|cc|cc|c}
        \toprule
        \multirow{2}{*}{\bf{Config}} & \multicolumn{2}{c|}{\bf{Loc. (mean Acc\%)}} & \multicolumn{2}{c|}{\bf{Fit. (mean \bm{$\rho$}})} & \multirow{2}{*}{\bf{Func. (mean $\mathbf{F}_\mathbf{max}$})} \\
        \cmidrule{2-3}
        \cmidrule{4-5}
        & Fix-enc. & Full-m. & Fix-enc. & Full-m. &  \\
        \midrule
        Full loss & 87.44 & 85.54 & 0.601 & 0.772 & 0.627 \\
        \midrule
        w/o $\mathcal{L}_{\mathrm{MPM}}\!$ & 87.40$_{(\downarrow\!\;0.05\%)}\!\!$ & $\!\!$85.12$_{(\downarrow\!\;0.49\%)}$ & 0.593$_{(\downarrow\!\;1.33\%)}\!\!$ & $\!\!$0.766$_{(\downarrow\!\;0.78\%)}$ & 0.625$_{(\downarrow\!\;0.32\%)}$ \\
        w/o $\mathcal{L}_{\mathrm{GC}}\!$ & 86.34$_{(\downarrow\!\;\textcolor{r1}{\mathbf{1.26\%}})}\!\!$ & $\!\!$85.21$_{(\downarrow\!\;0.39\%)}$ & 0.579$_{(\downarrow\!\;\textcolor{r1}{\mathbf{3.66\%}})}\!\!$ & $\!\!$0.758$_{(\downarrow\!\;1.81\%)}$ & 0.613$_{(\downarrow\!\;\textcolor{r1}{\mathbf{2.23\%}})}$ \\
        w/o $\mathcal{L}_{\mathrm{MMP}}\!$ & 87.41$_{(\downarrow\!\;0.03\%)}\!\!$ & $\!\!$84.97$_{(\downarrow\!\;\textcolor{r1}{\mathbf{0.67\%}})}$ & 0.588$_{(\downarrow\!\;2.16\%)}\!\!$ & $\!\!$0.751$_{(\downarrow\!\;\textcolor{r1}{\mathbf{2.72\%}})}$ & 0.615$_{(\downarrow\!\;1.91\%)}$ \\
        \bottomrule
    \end{tabular}
\end{adjustbox}
\end{spacing}
\vspace{-2mm}
\end{table}

%%%%%%%%%%%%%%%%%%%%%%%%%%%%%%%%%%%%%%%%%%%%%%%%%%%%%%%%%%%%

\subsection{Ablation Study} \label{sec:exp:ablation}

\textbf{Effect of Pre-training Data Source:} In this project, besides Swiss-Prot, we also tried to use TrEMBL~\cite{bairoch2000swiss} as the data source to construct ProtDescribe. Compared to Swiss-Prot with high-quality human annotations for around 500K proteins, TrEMBL contains a larger number of over 200M annotated proteins, while the TrEMBL annotations are given by computational tools and are thus less accurate and have lower protein property coverage (as shown in Tab.~\ref{tab:compare_coverage}). 

The results in Tab.~\ref{tab:compare_pretrain_source} show that the ProtST-ESM-1b pre-trained on the smaller while higher-quality Swiss-Prot-based dataset performs better. Therefore, for the multimodal pre-training of protein sequences and biomedical texts, data quality could be more important than data quantity. 

\textbf{Effect of Pre-training Losses:} Tab.~\ref{tab:ablation} reports the averaged performance of {\method}-ESM-1b by using full or partial pre-training losses (per-task results are in Appendix~\ref{supp:sec:ablation:loss}). By removing any of three pre-training losses, performance decay occurs on all three types of tasks. Such phenomenon verifies the necessity of each {\method} pre-training loss, where $\mathcal{L}_{\mathrm{GC}}$ and $\mathcal{L}_{\mathrm{MMP}}$ inject different granularities of protein property information into a PLM, and $\mathcal{L}_{\mathrm{MPM}}$ preserves the PLM's original representation power. 

\textbf{Effect of PLM:} According to the results in Tabs.~\ref{tab:loc-fit} and \ref{tab:anno}, we can observe that the strength of a {\method}-induced PLM correlates with the strength of its initial PLM. To be specific, the better performance of ESM-1b and ESM-2 over ProtBert is inherited by their {\method}-induced variants. 

%%%%%%%%%%%%%%%%%%%%%%%%%%%%%%%%%%%%%%%%%%%%%%%%%%%%%%%%%%%%

%%%%%%%%%%%%%%%%%%%%%%%%%%%%%%%%%%%%%%%%%%%%%%%%%%%%%%%%%%%%

\section{Related Work} \label{sec:rela}

\textbf{Protein Representation Learning:} Learning effective protein representations is of great importance for machine learning guided protein understanding. Existing works learn protein representations in two ways: (1) Sequence-based methods model protein sequences on evolutionary scale~\cite{elnaggar2020prottrans,rives2021biological,lin2022language} or on individual protein families~\cite{bileschi2019using,meier2021language,biswas2021low}; (2) Structure-based methods seek to represent different levels of protein structures including residue-level structures~\cite{gligorijevic2021structure,zhang2022protein,xu2022eurnet}, all-atom structures~\cite{jing2020learning,zhang2023physics} and protein surfaces~\cite{gainza2020deciphering,sverrisson2021fast}. Our work aims to enhance protein sequence representation learning by using textual protein property descriptions. 

%%%%%%%%%%%%%%%%%%%%%%%%%%%%%%%%%%%%%%%%%%%%%%%%%%%%%%%%%%%%

\textbf{Multimodal Representation Learning:} It has been broadly studied how to learn better image~\cite{radford2021learning,singh2022flava}, video~\cite{luo2020univl,xu2021videoclip}, speech~\cite{chung2020splat,qian2021speech} and molecule~\cite{edwards2021text2mol,liu2022multi} representations by incorporating text supervision, while such study is lacked for proteins. 
OntoProtein~\cite{zhang2022ontoprotein} learns protein representations under the context of a knowledge graph; ProGen~\cite{madani2020progen} incorporates protein function labels to generate functional proteins. 
However, these two works investigate less the effect of biomedical texts. 
% OntoProtein~\cite{zhang2022ontoprotein} makes relevant attempts along this direction while mainly by applying a knowledge graph instead of language. 
Our work takes the initiative of enhancing protein sequence representation learning by biomedical texts. 

%%%%%%%%%%%%%%%%%%%%%%%%%%%%%%%%%%%%%%%%%%%%%%%%%%%%%%%%%%%%

%%%%%%%%%%%%%%%%%%%%%%%%%%%%%%%%%%%%%%%%%%%%%%%%%%%%%%%%%%%%

\section{Conclusions and Future Work} \label{sec:conclusion}

In this work, we propose the {\method} framework to study how textual protein property descriptions can boost protein sequence pre-training and understanding. We build the {\dataset} dataset that aligns protein sequences with their diverse property descriptions. {\method} pre-training injects the property information with different granularities into a protein language model (PLM). The {\method}-induced PLMs are verified to be generally effective on various downstream applications including supervised learning, zero-shot protein classification and zero-shot text-to-protein retrieval. 

The current ProtDescribe dataset is limited in the coverage of protein sequences and textual property descriptions, which motivates us to resort to massive biomedical articles in PubMed~\cite{canese2013pubmed} for information extraction. 
In addition, we plan to extend the {\dataset} dataset by incorporating protein structures and study biomedical text enhanced protein structure representation learning. Also, we will go beyond text-to-protein retrieval towards text-guided controllable protein design. 

\section*{Acknowledgments}

The authors would like to thank Meng Qu, Zhaocheng Zhu, Zuobai Zhang and Hesham Mostafa for their helpful discussions and comments.

This project is supported by Intel-MILA partnership program, the Natural Sciences and Engineering Research Council (NSERC) Discovery Grant, the Canada CIFAR AI Chair Program, collaboration grants between Microsoft Research and Mila, Samsung Electronics Co., Ltd., Amazon Faculty Research Award, Tencent AI Lab Rhino-Bird Gift Fund, a NRC Collaborative R\&D Project (AI4D-CORE-06) as well as the IVADO Fundamental Research Project grant PRF-2019-3583139727.

%%%%%%%%%%%%%%%%%%%%%%%%%%%%%%%%%%%%%%%%%%%%%%%%%%%%%%%%%%%%

\newpage
\bibliography{references}
\bibliographystyle{icml2023}


% \newpage
\newpage
\appendix

%%%%%%%%%%%%%%%%%%%%%%%%%%%%%%%%%%%%%%%%%%%%%%%%%%%%%%%%%%%%

\section{Model Architecture for Pre-training}  \label{supp:sec:pretrain_arch}

\begin{figure}[t]
\centering
    \includegraphics[width=1.0\linewidth]{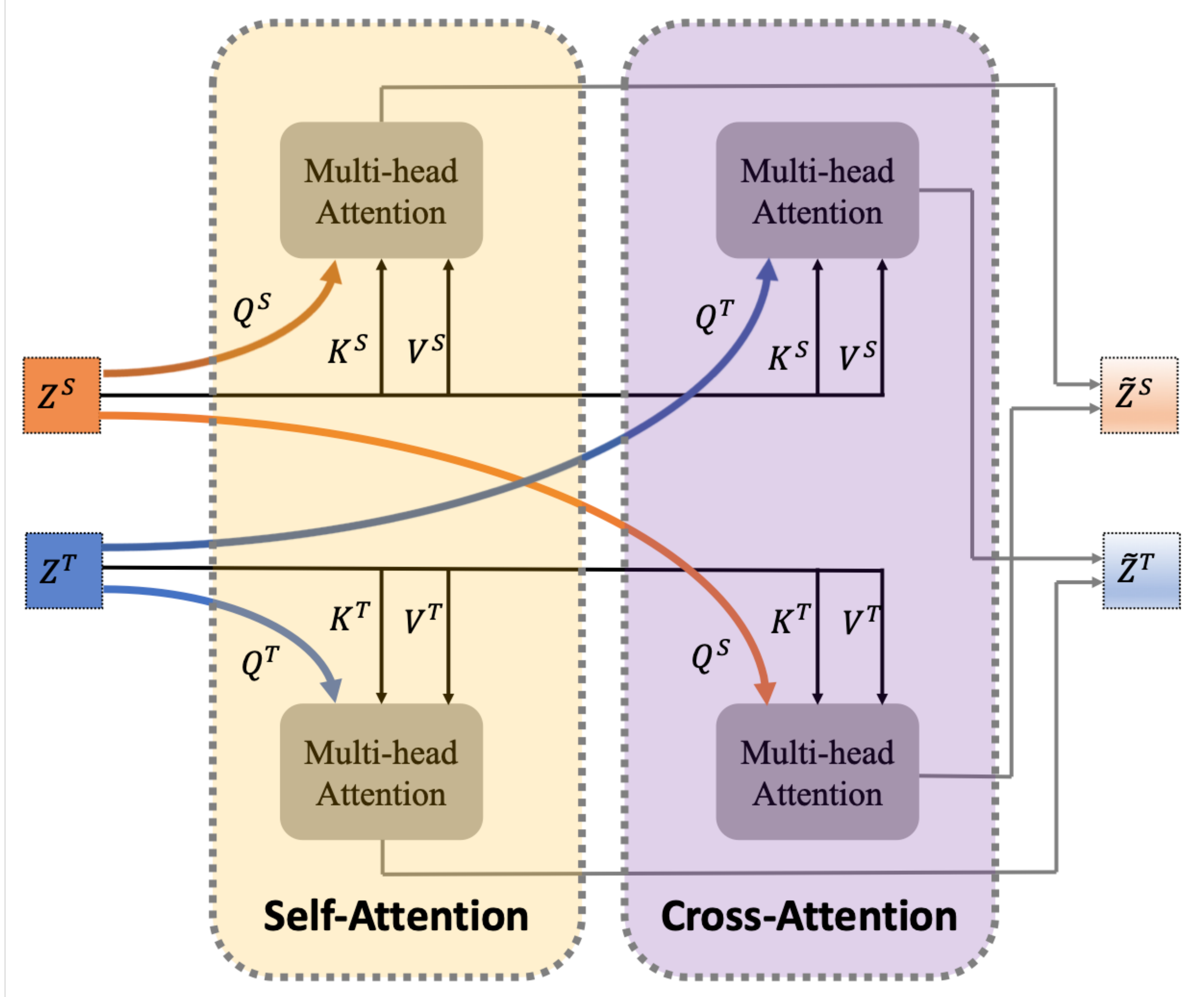}
    \vspace{-6.5mm}
    \caption{\textbf{Architecture of the fusion layer.} This layer fuses the protein representation and the text representation by querying over them with self-attention and cross-attention.
    }
    \label{fig:fusion_model}
    \vspace{-3mm}
\end{figure}

\textbf{Fusion Module:} The fusion module extracts multimodal representations from the unimodal representations of protein sequence and text description. As shown in Fig.~\ref{fig:fusion_model}, each \emph{fusion layer} of this module receives a sequence of residue representations $Z^S = [z^s_1, z^s_2, \cdots , z^s_n] \in \mathbb{R}^{n \times d}$ and a sequence of word representations $Z^T = [z^t_1, z^t_2, \cdots , z^t_m] \in \mathbb{R}^{m \times d}$ ($d$ denotes the hidden dimension), and the layer updates each residue/word representation by attending to all residues and all words. Specifically, two sets of projection matrices $(W_q^S, W_k^S, W_v^S)$ and $(W_q^T, W_k^T, W_v^T)$ are respectively used to derive the queries, keys and values for protein sequence and text description as below (each projection matrix is in $\mathbb{R}^{d \times d}$):
\begin{equation} \label{supp:eq:residue_qkv}
    Q^S = Z^S W_q^S, \quad K^S = Z^S W_k^S, \quad V^S = Z^S W_v^S, 
\end{equation}
\begin{equation} \label{supp:eq:word_qkv}
    Q^T = Z^T W_q^T, \quad K^T = Z^T W_k^T, \quad V^T = Z^T W_v^T, 
\end{equation}
where $Q^S, K^S, V^S \in \mathbb{R}^{n \times d}$ are the queries, keys and values for protein sequence, and $Q^T, K^T, V^T \in \mathbb{R}^{m \times d}$ are the queries, keys and values for text description. Multi-head self- and cross-attention are then applied to update each residue and word representation as below: 
\begin{equation} \label{supp:eq:seq_attn}
\small
\tilde{Z}^S = \frac{1}{2} \big ( \mathrm{MHA} (Q^S, K^S, V^S) + \mathrm{MHA} (Q^S, K^T, V^T) \big) ,
\end{equation}
\begin{equation} \label{supp:eq:text_attn}
\small
\tilde{Z}^T = \frac{1}{2} \big ( \mathrm{MHA} (Q^T, K^T, V^T) + \mathrm{MHA} (Q^T, K^S, V^S) \big) ,
\end{equation}
where $\tilde{Z}^S \in \mathbb{R}^{n \times d}$ and $\tilde{Z}^T \in \mathbb{R}^{m \times d}$ are the updated residue and word representations, and $\mathrm{MHA}(\cdot,\cdot,\cdot)$ denotes the multi-head attention operation~\cite{vaswani2017attention}. 

In our implementation, each fusion layer contains 8 attention heads, and we equip the fusion module with a single fusion layer so as to restrict the capacity of fusion module and facilitate the representation power of PLM. Upon the fused residue and word representations produced by the fusion module, multimodal mask prediction is performed. 

%%%%%%%%%%%%%%%%%%%%%%%%%%%%%%%%%%%%%%%%%%%%%%%%%%%%%%%%%%%%

%%%%%%%%%%%%%%%%%%%%%%%%%%%%%%%%%%%%%%%%%%%%%%%%%%%%%%%%%%%%

\textbf{Projection Head for Multimodal Representation Alignment:} Following SimCLR~\cite{chen2020simple}, we use a two-layer MLP (with ReLU nonlinearity in between) to project the protein sequence representation extracted by the PLM, and another two-layer nonlinear MLP is employed to project the text description representation extracted by the BLM. The projected sequence and text representations are then used to compute the global contrastive loss defined in Eq.~(\ref{eq:infonce}). 

\textbf{Prediction Head for Masked Protein Modeling (MPM):} Based on the residue representations extracted by the PLM, we utilize a two-layer MLP (with ReLU nonlinearity in between) to predict the type of each residue token masked at input. 

\textbf{Prediction Head for Multimodal Mask Prediction (MMP):} Upon the fused residue representations output from the fusion module, a two-layer MLP (with ReLU nonlinearity in between) is used to predict the type of each residue token masked at input protein sequence. Upon the fused word representations produced by the fusion module, another two-layer nonlinear MLP is employed to predict each word token masked at input text description. 

%%%%%%%%%%%%%%%%%%%%%%%%%%%%%%%%%%%%%%%%%%%%%%%%%%%%%%%%%%%%

%%%%%%%%%%%%%%%%%%%%%%%%%%%%%%%%%%%%%%%%%%%%%%%%%%%%%%%%%%%%

\section{More Experimental Setups} \label{supp:sec:setup}

%%%%%%%%%%%%%%%%%%%%%%%%%%%%%%%%%%%%%%%%%%%%%%%%%%%%%%%%%%%%

\begin{table*}[t]
\begin{spacing}{1.05}
\centering
\caption{Examples of property descriptions in the {\dataset} dataset. We index each description with the Swiss-Prot entry name of its corresponding protein.}
\vspace{-2.5mm}
\label{supp:tab:descr_example}
\begin{adjustbox}{max width=1\linewidth}
    \begin{tabular}{p{0.12\linewidth}|p{0.88\linewidth}}
        \toprule
        \bf{Entry name$\;\;\;$} & \bf{Description} \\
        \midrule
        14336\_ORYSJ & \textcolor{highlight}{PROTEIN NAME}: 14-3-3-like protein GF14-F. \textcolor{highlight}{FUNCTION}: Is associated with a DNA binding complex that binds to the G box, a well-characterized cis-acting DNA regulatory element found in plant genes. \textcolor{highlight}{SUBCELLULAR LOCATION}: Cytoplasm. Nucleus. \textcolor{highlight}{SIMILARITY}: Belongs to the 14-3-3 family. \\
        \midrule
        053R\_FRG3G & \textcolor{highlight}{PROTEIN NAME}: Putative myristoylated protein 053R. \textcolor{highlight}{FUNCTION}: May play a critical role in virion formation. Essential for virus replication in vitro. \textcolor{highlight}{SUBCELLULAR LOCATION}: Host membrane; Multi-pass membrane protein. \\
        \midrule
        1A16\_ORYSJ & \textcolor{highlight}{PROTEIN NAME}: 1-aminocyclopropane-1-carboxylate synthase 6. \textcolor{highlight}{FUNCTION}: Catalyzes the formation of 1-aminocyclopropane-1-carboxylate, a direct precursor of ethylene in higher plants (By similarity). Required for the regulation of starch grain size in endosperm. \textcolor{highlight}{SUBCELLULAR LOCATION}: Plastid, amyloplast membrane. Note=Localizes to the amyloplast membrane surrounding starch grains in endosperm, pollen, and pericarp. \textcolor{highlight}{SIMILARITY}: Belongs to the class-I pyridoxal-phosphate-dependent aminotransferase family. \\
        \midrule
        17KD\_RICPR & \textcolor{highlight}{PROTEIN NAME}: 17 kDa surface antigen. \textcolor{highlight}{SUBCELLULAR LOCATION}: Cell outer membrane; Lipid-anchor. \textcolor{highlight}{SIMILARITY}: Belongs to the rickettsiale 17 kDa surface antigen family. \\
        \midrule
        1A1D\_CYBSA & \textcolor{highlight}{PROTEIN NAME}: 1-aminocyclopropane-1-carboxylate deaminase. \textcolor{highlight}{FUNCTION}: Catalyzes a cyclopropane ring-opening reaction, the irreversible conversion of 1-aminocyclopropane-1-carboxylate (ACC) to ammonia and alpha-ketobutyrate. \textcolor{highlight}{SIMILARITY}: Belongs to the ACC deaminase/D-cysteine desulfhydrase family. \\
        \midrule
        1AP1\_BRAOT & \textcolor{highlight}{PROTEIN NAME}: Floral homeotic protein APETALA 1-1. \textcolor{highlight}{FUNCTION}: Transcription factor that promotes early floral meristem identity in synergy with LEAFY. Displays a redundant function with CAULIFLOWER in the up-regulation of LEAFY. Required subsequently for the transition of an inflorescence meristem into a floral meristem, and for the normal development of sepals and petals in flowers. Regulates positively B class homeotic proteins (By similarity). \textcolor{highlight}{SUBCELLULAR LOCATION}: Nucleus. \\
        \bottomrule
    \end{tabular}
\end{adjustbox}
\end{spacing}
\end{table*}

%%%%%%%%%%%%%%%%%%%%%%%%%%%%%%%%%%%%%%%%%%%%%%%%%%%%%%%%%%%%

\begin{table}[t]
\begin{spacing}{1.05}
\centering
\caption{{\method} pre-training configurations. \emph{Abbr.}, lr.: learning rate; bs.: batch size.}
\vspace{-2.5mm}
\label{supp:tab:pretrain_config}
\begin{adjustbox}{max width=1\linewidth}
    \begin{tabular}{l|ccccc}
        \toprule
        \bf{Model} & \bf{optimizer} & \bf{lr.} & \bf{bs.} & \bf{\#epochs} & \bf{train time} \\
        \midrule
        \bf{{\method}-ProtBert} & Adam & $1.0 \times 10^{-5}$ & 16 & 20 & 117h 10min \\
        \bf{{\method}-ESM-1b} & Adam & $1.0 \times 10^{-5}$ & 12 & 20 & 205h 36min \\
        \bf{{\method}-ESM-2} & Adam & $1.0 \times 10^{-5}$ & 12 & 20 & 206h 12min \\
        \bottomrule
    \end{tabular}
\end{adjustbox}
\end{spacing}
\end{table}

%%%%%%%%%%%%%%%%%%%%%%%%%%%%%%%%%%%%%%%%%%%%%%%%%%%%%%%%%%%%

\subsection{More Pre-training Setups} \label{supp:sec:setup:pretrain}

\textbf{Pre-training Data Curation:} We add prefixes to denote annotations from different fields, \emph{i.e.}, ``PROTEIN NAME'' for the protein name field, ``FUNCTION'' for the protein function field, ``SUBCELLULAR LOCATION'' for the subcellular location field, and ``SIMILARITY'' for the protein family field. The complete protein property description is formed by concatenating all annotations of the protein in the order of (1) protein name, (2) protein function, (3) subcellular location, and (4) protein family. In Tab.~\ref{supp:tab:descr_example}, we present several property descriptions coupled with the Swiss-Prot entry names of their corresponding proteins. 

\textbf{Training Configurations:} We list the training configurations of three {\method}-induced PLMs in Tab.~\ref{supp:tab:pretrain_config}. In general, an Adam optimizer with the constant learning rate of $1.0 \times 10^{-5}$ is used to train the model for 20 epochs on 4 Tesla V100 GPUs, where {\method}-ProtBert adopts the batch size of 16 (4 proteins per GPU), and {\method}-ESM-1b and {\method}-ESM-2 adopt the batch size of 12 (3 proteins per GPU). Since the PLM is pre-trained, we set its learning rate as $1.0 \times 10^{-6}$, \emph{i.e.}, one tenth of other modules. The weights of PubMedBERT are frozen along the whole process. To reduce the memory cost, we truncate the protein sequences that have more than 450 residues to the length of 450, where the truncation starts from a random residue before the last 450 ones. Following MoCo~\cite{he2020momentum}, we initialize the temperature parameter $\tau$ in Eq.~(\ref{eq:infonce}) as 0.07 and optimize it along the training process. 

%%%%%%%%%%%%%%%%%%%%%%%%%%%%%%%%%%%%%%%%%%%%%%%%%%%%%%%%%%%%

\begin{table}[t]
\begin{spacing}{1.05}
\centering
\caption{Configurations of fix-encoder learning and full-model tuning on three task types. \emph{Abbr.}, lr.: learning rate; bs.: batch size; MSE: mean squared error; CE: cross entropy; BCE: binary cross entropy.}
\vspace{-2.5mm}
\label{supp:tab:repr_config}
\begin{adjustbox}{max width=1\linewidth}
    \begin{tabular}{l|ccccc}
        \toprule
        \bf{Task} & \bf{optimizer} & \bf{lr.} & \bf{bs.} & \bf{\#epochs} & \bf{loss} \\
        \midrule
        \multicolumn{6}{c}{\bf{fix-encoder learning}} \\
        \midrule
        \bf{Localization} & Adam & $5.0 \times 10^{-5}$ & 128 & 100 & CE \\
        \bf{Fitness} & Adam & $5.0 \times 10^{-5}$ & 128 & 100 & MSE \\
        \midrule
        \multicolumn{6}{c}{\bf{full-model tuning}} \\
        \midrule
        \bf{Localization} & Adam & $2.0 \times 10^{-4}$ & 12 & 100 & CE \\
        \bf{Fitness} & Adam & $2.0 \times 10^{-4}$ & 24 & 100 & MSE \\
        \bf{Annotation} & Adam & $1.0 \times 10^{-4}$ & 8 & 50 & BCE \\
        \bottomrule
    \end{tabular}
\end{adjustbox}
\end{spacing}
\end{table}

%%%%%%%%%%%%%%%%%%%%%%%%%%%%%%%%%%%%%%%%%%%%%%%%%%%%%%%%%%%%

\subsection{More Representation Learning Setups} \label{supp:sec:setup:repr}

\textbf{Architecture of Prediction Heads:} Following the default settings in TorchDrug~\cite{zhu2022torchdrug}, the prediction of each task is performed by a two-layer MLP with ReLU nonlinearity in between. To be specific, given the protein representation, the MLP head is used to predict classification logits for localization prediction, regression score for fitness prediction and per-function classification logits for function annotation. 

\textbf{Training Configurations:} In Tab.~\ref{supp:tab:repr_config}, we present the detailed configurations of fix-encoder learning and full-model tuning on three task types, which mainly follows the configurations used in PEER benchmark~\cite{xu2022peer}. For full-model tuning, the learning rate of the PLM is set as one tenth of the value in Tab.~\ref{supp:tab:repr_config}. The protein sequence encoders trained from scratch do not use smaller learning rates. All experiments are conducted on 4 Tesla V100 GPUs. 

\textbf{Evaluation Metrics:} The protein function annotation tasks are measured by AUPR and $\mathrm{F}_{\mathrm{max}}$. We clarify their definitions as below:

(1) \textbf{AUPR} denotes the pair-centric area under precision-recall curve. It computes the average precision scores for all protein-function pairs, which is exactly the micro-average precision score for the multiple binary classification problem. 

(2) $\mathbf{F}_{\mathbf{max}}$ denotes the protein-centric maximum F-score. Given a decision threshold $t \in [0,1]$, it first calculates the precision and recall for each protein:
\begin{equation} \label{supp:eq:fmax:precision}
    \text{precision}_i(t)=\frac{\sum_f \mathbbm{1}[f\in P_i(t) \cap T_i]}{\sum_f \mathbbm{1}[f\in P_i(t)]}, 
\end{equation}
\begin{equation} \label{supp:eq:fmax:recall}
    \text{recall}_i(t)=\frac{\sum_f \mathbbm{1}[f\in P_i(t) \cap T_i]}{\sum_f \mathbbm{1}[f\in T_i]},
\end{equation}
where $f$ denotes a functional term of EC or GO, $T_i$ is the set collecting all experimentally determined functions for protein $i$, $P_i(t)$ denotes the predicted functions for protein $i$ whose scores are at least $t$, and $\mathbbm{1}[\cdot]$ represents the indicator function. The precision and recall are then averaged over all proteins:
\begin{equation} \label{supp:eq:fmax:avg_precision}
    \text{precision}(t)=\frac{1}{M(t)}\sum_{i} \text{precision}_i(t),
\end{equation}
\begin{equation} \label{supp:eq:fmax:avg_recall}
    \text{recall}(t)=\frac{1}{N}\sum_{i} \text{recall}_i(t),
\end{equation}
where $N$ is the total number of proteins, and $M(t)$ denotes the number of proteins that contain at least one prediction larger than $t$, \emph{i.e.}, $|P_i(t)| > 0$. 

Finally, the $\mathrm{F}_{\mathrm{max}}$ score is computed as the maximum value of F-measure over all thresholds:
\begin{equation} \label{supp:eq:fmax}
    \mathrm{F}_{\mathrm{max}}=
    \max_t\left\{\frac{2\cdot \text{precision}(t)\cdot \text{recall}(t)}{\text{precision}(t)+ \text{recall}(t)}\right\}.
\end{equation}

%%%%%%%%%%%%%%%%%%%%%%%%%%%%%%%%%%%%%%%%%%%%%%%%%%%%%%%%%%%%

\begin{table}[t]
\begin{spacing}{1.05}
\centering
\caption{\small Zero-shot protein classification performance under different prompt templates. \emph{Abbr.}, Acc: accuracy; loc.: localization.}
\vspace{-2.5mm}
\label{supp:tab:prompt}
\begin{adjustbox}{max width=1.0\linewidth}
    \begin{tabular}{ll|cc}
        \toprule
        \bf{Prompt template} & \bf{Label} & \bf{Subcellular loc. (\emph{Acc\%})} & \bf{Reaction (\emph{Acc\%})} \\
        \midrule
        Name only & Name & 25.68 & 25.27 \\
        Natural language & Name & 36.24 & 26.93 \\
        \bf{Pre-training template} & \bf{Name} & \bf{43.49} & \bf{29.85} \\
        Pre-training template & Description & 29.90 & 21.91 \\
        \bottomrule
    \end{tabular}
\end{adjustbox}
\end{spacing}
\end{table}

%%%%%%%%%%%%%%%%%%%%%%%%%%%%%%%%%%%%%%%%%%%%%%%%%%%%%%%%%%%%

\subsection{More Zero-shot Protein Classification Setups} \label{supp:sec:setup:zero}

\textbf{Prompt Engineering for Subcellular Localization Prediction:} Based on the information provided by DeepLoc~\cite{almagro2017deeploc}, we consider two label formats, the \emph{name} of each subcellular location (\emph{i.e.}, the ``Location'' field in the Tab.~1 of DeepLoc paper) and the \emph{description} of each location (\emph{i.e.}, the ``Sublocations'' field in the Tab.~1 of DeepLoc paper). We further embed the labels into three prompt templates: (1) \emph{Name only}: only the label itself is used; (2) \emph{Natural language}: the label is embedded into the template ``A protein locating at \{\texttt{label}\}.''; (3) \emph{Pre-training template}: the label is embedded into the template ``SUBCELLULAR LOCATION: \{\texttt{label}\}''. 

According to the results in Tab.~\ref{supp:tab:prompt}, we can observe that the pre-training template clearly outperforms other two templates on the subcellular localization prediction task, which mainly owes to the alignment of text format across pre-training and zero-shot prediction. It is shown that representing the labels with location names leads to better performance than using location descriptions, since the location names better fit the biomedical text distribution that the BLM is trained on. Based on these results, we represent the labels with the location names coupled with the pre-training prompt template on this task. 

\textbf{Prompt Engineering for Reaction Classification:} Same as subcellular localization prediction, we also use two sets of label notations for reaction classification, \emph{i.e.}, the \emph{name} and the \emph{description}. (1) The \emph{name} refers to the composition of the enzyme class name and its alternative names, allowing unambiguous identification of each enzyme class. (2) The \emph{description} further adds the scientific comments that discuss each class of enzymes in depth, which are extracted from scientific articles published by the International Union of Biochemistry and Molecular Biology (IUBMB). We retrieve all the information from \citet{chang2021brenda}. 

We embed such label information into three prompt templates: (1) \emph{Name only}: the concatenation of the name and alternative names of an enzyme class, \emph{i.e.}, ``\{\texttt{Name}\} \{\texttt{AlterNames}\}''; (2) \emph{Natural Language}: the label is incorporated into a natural-language-like template ``A \{\texttt{Name}\} enzyme. This enzyme is also known as \{\texttt{AlterNames}\}.''; (3) \emph{Pre-training template}: the label is merged into the template used for pre-training, \emph{i.e.}, ``FUNCTION: \{\texttt{Name}\} \{\texttt{AlterNames}\}'' (scientific comments ``\{\texttt{Comments}\}'' are appended after the names if the \emph{description} is used). 

% We also embed the two labels into three prompt templates: (1) \emph{Name only}: the two sub-fields in the label are concatenated to form the template ``\{\texttt{label.EnzymeName}\} \{\texttt{label.AlterNames}\}''; (2) \emph{Natural Language}: the label is incorporated into the natural-language-like template ``A \{\texttt{label.EnzymeName}\} enzyme. This enzyme is also known as \{\texttt{label.AlterNames}\}''; (3) \emph{Pre-training template}: the label is incorporated into the template ``FUNCTION: \{\texttt{label.EnzymeName}\} \{\texttt{label.AlterNames}\}'', following the same usage of property fields during pretraining.

According to Tab.~\ref{supp:tab:prompt},
the pre-training template performs the best on the reaction classification task, mainly thanks to the consistent format of text descriptions between pre-training and zero-shot prediction. 
Injecting detailed scientific comments does not bring further benefits to the zero-shot performance. 
% To our surprise, injecting detailed scientific comments even decreases the performance, which we leave to future work to determine the root causes. 
Therefore, we represent each enzyme class with its name and alternative names along with the pre-training prompt template for this task. 

\textbf{Nonparametric Few-shot Classifier:} We adopt the nonparametric classifier proposed by \citet{khandelwal2019generalization} as baseline. Specifically, given $n$-shot $K$-class training samples $\{ \{(S^k_i, y^k_i=k)\}_{i=1}^n \}_{k=1}^K$ composed of pairs of protein sequence and label, we employ the PLM to extract the representations $\{ \{z^k_i\}_{i=1}^n \}_{k=1}^K$ of all protein sequences. When a test protein $S'$ comes, the nonparametric classifier first extracts its representation $z'$ via the PLM and then derives its classification logits $\{y'_k\}_{k=1}^K$ by computing its representation similarity with each training protein:
\begin{equation} \label{supp:eq:rbf}
y'_k = \sum_{i=1}^n \exp \big ( \! - \! || z' - z^k_i ||^2_2 \big), \quad k=1, \cdots, K .
\end{equation}
Softmax is performed upon these logits to derive classification probabilities. Such a classifier predicts based on the relations between test sample and training samples, which well fits the few-shot setting. In our experiments, the nonparametric classifier based on ESM-1b and the one based on {\method}-ESM-1b serve as two baselines for zero-shot classifiers. 

%%%%%%%%%%%%%%%%%%%%%%%%%%%%%%%%%%%%%%%%%%%%%%%%%%%%%%%%%%%%

%%%%%%%%%%%%%%%%%%%%%%%%%%%%%%%%%%%%%%%%%%%%%%%%%%%%%%%%%%%%

\begin{table}[t]
\begin{spacing}{1.05}
\centering
\caption{Performance comparison of PLMs on ProteinGym Substitution benchmark. \emph{Abbr.}, retr.: retrieval.}
\vspace{-2mm}
\label{supp:tab:proteingym:plm}
\begin{adjustbox}{max width=1\linewidth}
    \begin{tabular}{l|ccccc}
        \toprule
        \bf{Model} & {\method}-ESM-1b & ESM-1b & ESM-1v & Tranception L (\emph{w/o} retr.) & Progen2 XL \\
        \midrule
        \bf{Model Type} & PLM & PLM & PLM & PLM & PLM \\
        \midrule
        \midrule
        \bf{UniProt-level Mean $\rho$} & \bf{0.412} & 0.358 & 0.372 & 0.401 & 0.402 \\
        \bottomrule
    \end{tabular}
\end{adjustbox}
\end{spacing}
\vspace{-2mm}
\end{table}

%%%%%%%%%%%%%%%%%%%%%%%%%%%%%%%%%%%%%%%%%%%%%%%%%%%%%%%%%%%%

\begin{table}[t]
\begin{spacing}{1.05}
\centering
\caption{{\method}-ESM-1b \emph{v.s.} alignment-based methods on ProteinGym Substitution benchmark.}
\vspace{-2mm}
\label{supp:tab:proteingym:compare}
\begin{adjustbox}{max width=1\linewidth}
    \begin{tabular}{l|cccc}
        \toprule
        \bf{Model} & {\method}-ESM-1b & EVE & GEMME & {\method}-ESM-1b + GEMME \\
        \midrule
        \bf{Model Type} & PLM & Align & Align & Hybrid \\
        \midrule
        \midrule
        \bf{UniProt-level Mean $\rho$} & 0.412 & 0.443 & 0.459 & \bf{0.464} \\
        \bottomrule
    \end{tabular}
\end{adjustbox}
\end{spacing}
\end{table}

%%%%%%%%%%%%%%%%%%%%%%%%%%%%%%%%%%%%%%%%%%%%%%%%%%%%%%%%%%%%

\section{Experimental Results on ProteinGym}  \label{supp:sec:proteingym}

%%%%%%%%%%%%%%%%%%%%%%%%%%%%%%%%%%%%%%%%%%%%%%%%%%%%%%%%%%%%

\subsection{Comparisons of Protein Language Models (PLMs)}  \label{supp:sec:proteingym:plm}

\textbf{Baselines.} We compare the proposed {\method}-ESM-1b with four performant PLMs, \emph{i.e.}, ESM-1b~\cite{rives2021biological}, ESM-1v~\cite{meier2021language}, Tranception L (\emph{w/o} retrieval)~\cite{notin2022tranception} and Progen2 XL~\cite{nijkamp2022progen2}. Note that, for fair comparison, we do not include the PLMs with model ensemble (\emph{e.g.}, VESPA~\cite{marquet2022embeddings}) and the PLMs with inference-time retrieval (\emph{e.g.}, Tranception L w/ retrieval~\cite{notin2022tranception}). We report the UniProt-level Mean Spearman’s $\rho$. 

\textbf{Results.} Under such a fair comparison, in Tab.~\ref{supp:tab:proteingym:plm}, {\method}-ESM-1b achieves the best performance. In particular, compared with ESM-1b (\emph{i.e.}, the initial PLM that {\method}-ESM-1b is based on), {\method}-ESM-1b obtains a significant performance gain with 15.1\% relative improvement. This result demonstrates the effectiveness of the proposed multimodal training, which injects protein property knowledge into the ESM-1b and enhances its downstream fitness prediction performance.

%%%%%%%%%%%%%%%%%%%%%%%%%%%%%%%%%%%%%%%%%%%%%%%%%%%%%%%%%%%%

\subsection{Comparisons with Alignment-based Methods}  \label{supp:sec:proteingym:compare}

\textbf{Baselines.} In this experiment, we involve two alignment-based methods, \emph{i.e.}, EVE~\cite{frazer2021disease} and GEMME~\cite{laine2019gemme}, for comparison. We further investigate the ensemble of {\method}-ESM-1b and GEMME. We report the UniProt-level Mean Spearman’s $\rho$. 

\textbf{Results.} In Tab.~\ref{supp:tab:proteingym:compare}, it is observed that the alignment-based methods are superior over {\method}-ESM-1b, since they additionally utilize the homologous information within sequence alignments, which is not utilized by {\method}-ESM-1b. However, by combining the normalized predictions of {\method}-ESM-1b and GEMME, the ensemble model ``{\method}-ESM-1b + GEMME'' outperforms these two SOTA alignment-based methods. This result verifies the complementary knowledge hidden in {\method}-ESM-1b and an alignment-based model in terms of fitness prediction. Therefore, it will be a promising direction to study the combination of these two lines of methods. We leave this as our future work.

%%%%%%%%%%%%%%%%%%%%%%%%%%%%%%%%%%%%%%%%%%%%%%%%%%%%%%%%%%%%

%%%%%%%%%%%%%%%%%%%%%%%%%%%%%%%%%%%%%%%%%%%%%%%%%%%%%%%%%%%%

\begin{table}[t]
\begin{spacing}{1.05}
\centering
\caption{\small Ablation study of pre-training losses on localization and fitness prediction. \emph{Abbr.}, Loc.: Localization; pred.: prediction; Acc: accuracy. \textcolor{gray}{Gray} denotes the performance decay.}
\vspace{-2.5mm}
\label{supp:tab:ablation:loss:loc-fit}
\begin{adjustbox}{max width=1\linewidth}
% \begin{threeparttable}
    \begin{tabular}{l|cc|ccccc|c}
        \toprule
        \multirow{2}{*}{\bf{Model}} & \multicolumn{2}{c|}{\bf{Loc. pred.} (\emph{Acc\%})} &
        \multicolumn{6}{c}{\bf{Fitness pred.} (\emph{Spearman's \bm{$\rho$}})} \\
        \cmidrule{2-3}
        \cmidrule{4-9}
        & \bf{Bin} & \bf{Sub} & \bf{$\beta$-lac} & \bf{AAV} & \bf{Thermo} & \bf{Flu} & \bf{Sta} & \bf{Mean \bm{$\rho$}} \\
        \midrule
        \multicolumn{9}{c}{\bf{Fix-encoder learning}} \\
        \midrule
        {\method}-ESM-1b & 92.87 & 82.00 & 0.578 & 0.460 & 0.680 & 0.523 & 0.766 & 0.601 \\
        \midrule
        {\method}-ESM-1b (w/o $\mathcal{L}_{\mathrm{MPM}}$) & \cellcolor{decay} 92.52 & 82.28 & \cellcolor{decay} 0.558 & 0.475 & 0.680 & \cellcolor{decay} 0.522 & \cellcolor{decay} 0.730 & \cellcolor{decay} 0.593 \\
        {\method}-ESM-1b (w/o $\mathcal{L}_{\mathrm{GC}}$) & \cellcolor{decay} 92.12 & \cellcolor{decay} 80.55 & \cellcolor{decay} 0.560 & \cellcolor{decay} 0.448 & 0.684 & \cellcolor{decay} 0.467 & \cellcolor{decay} 0.738 & \cellcolor{decay} 0.579 \\
        {\method}-ESM-1b (w/o $\mathcal{L}_{\mathrm{MMP}}$) & \cellcolor{decay} 92.81 & 82.00 & \cellcolor{decay} 0.544 & 0.479 & 0.681 & \cellcolor{decay} 0.504 & \cellcolor{decay} 0.731 & \cellcolor{decay} 0.588 \\
        \midrule
        \multicolumn{9}{c}{\bf{Full-model tuning}} \\
        \midrule
        {\method}-ESM-1b & 92.35 & 78.73 & 0.895 & 0.850 & 0.681 & 0.682 & 0.751 & 0.772 \\
        \midrule
        {\method}-ESM-1b (w/o $\mathcal{L}_{\mathrm{MPM}}$) & 92.64 & \cellcolor{decay} 77.59 & \cellcolor{decay} 0.894 & \cellcolor{decay} 0.842 & 0.681 & 0.685 & \cellcolor{decay} 0.726 & \cellcolor{decay} 0.766 \\
        {\method}-ESM-1b (w/o $\mathcal{L}_{\mathrm{GC}}$) & \cellcolor{decay} 91.67 & 78.75 & \cellcolor{decay} 0.891 & \cellcolor{decay} 0.798 & \cellcolor{decay} 0.674 & 0.686 & \cellcolor{decay} 0.741 & \cellcolor{decay} 0.758 \\
        {\method}-ESM-1b (w/o $\mathcal{L}_{\mathrm{MMP}}$) & \cellcolor{decay} 91.90 & \cellcolor{decay} 78.03 & 0.902 & \cellcolor{decay} 0.804 & \cellcolor{decay} 0.677 & \cellcolor{decay} 0.678 & \cellcolor{decay} 0.696 & \cellcolor{decay} 0.751 \\
        \bottomrule
    \end{tabular}
% \begin{tablenotes}
%     \item[*] Fix-encoder learning: the PLM is used as a feature extractor with pre-trained weights frozen.
% \end{tablenotes}
% \end{threeparttable}
\end{adjustbox}
\end{spacing}
\vspace{-2mm}
\end{table}

%%%%%%%%%%%%%%%%%%%%%%%%%%%%%%%%%%%%%%%%%%%%%%%%%%%%%%%%%%%%

\begin{table}[t]
\begin{spacing}{1.07}
\centering
\caption{\small Ablation study of pre-training losses on function annotation. \textcolor{gray}{Gray} denotes the performance decay.}
\vspace{-2.5mm}
\label{supp:tab:ablation:loss:anno}
\begin{adjustbox}{max width=1\linewidth}
    \begin{tabular}{l|cc|cc|cc|cc}
        \toprule
        \multirow{2}{*}{\bf{Model}} & \multicolumn{2}{c|}{\bf{EC}} & \multicolumn{2}{c|}{\bf{GO-BP}} & \multicolumn{2}{c|}{\bf{GO-MF}} & \multicolumn{2}{c}{\bf{GO-CC}} \\
        \cmidrule{2-3}
        \cmidrule{4-5}
        \cmidrule{6-7}
        \cmidrule{8-9}
        & AUPR & $\mathrm{F}_{\mathrm{max}}$ & AUPR & $\mathrm{F}_{\mathrm{max}}$ & AUPR & $\mathrm{F}_{\mathrm{max}}$ & AUPR & $\mathrm{F}_{\mathrm{max}}$ \\
        \midrule
        \multicolumn{9}{c}{\bf{Full-model tuning}} \\
        \midrule
        {\method}-ESM-1b & 0.894 & 0.878 & 0.328 & 0.480 & 0.644 & 0.661 & 0.364 & 0.488 \\
        \midrule
        {\method}-ESM-1b (w/o $\mathcal{L}_{\mathrm{MPM}}$) & 0.898 & \cellcolor{decay} 0.873 & \cellcolor{decay} 0.324 & 0.483 & \cellcolor{decay} 0.642 & \cellcolor{decay} 0.660 & \cellcolor{decay} 0.350 & \cellcolor{decay} 0.482 \\
        {\method}-ESM-1b (w/o $\mathcal{L}_{\mathrm{GC}}$) & 0.894 & \cellcolor{decay} 0.870 & \cellcolor{decay} 0.322 & \cellcolor{decay} 0.463 & \cellcolor{decay} 0.638 & \cellcolor{decay} 0.656 & \cellcolor{decay} 0.327 & \cellcolor{decay} 0.462 \\
        {\method}-ESM-1b (w/o $\mathcal{L}_{\mathrm{MMP}}$) & \cellcolor{decay} 0.890 & \cellcolor{decay} 0.871 & 0.328 & \cellcolor{decay} 0.456 & \cellcolor{decay} 0.635 & \cellcolor{decay} 0.659 & \cellcolor{decay} 0.340 & \cellcolor{decay} 0.473 \\
        \bottomrule
    \end{tabular}
\end{adjustbox}
\end{spacing}
\end{table}

%%%%%%%%%%%%%%%%%%%%%%%%%%%%%%%%%%%%%%%%%%%%%%%%%%%%%%%%%%%%

\section{More Zero-shot Text-to-Protein Retrieval Results}  \label{supp:sec:t2p}

\begin{figure*}[t]
\centering
    \includegraphics[width=1.0\linewidth]{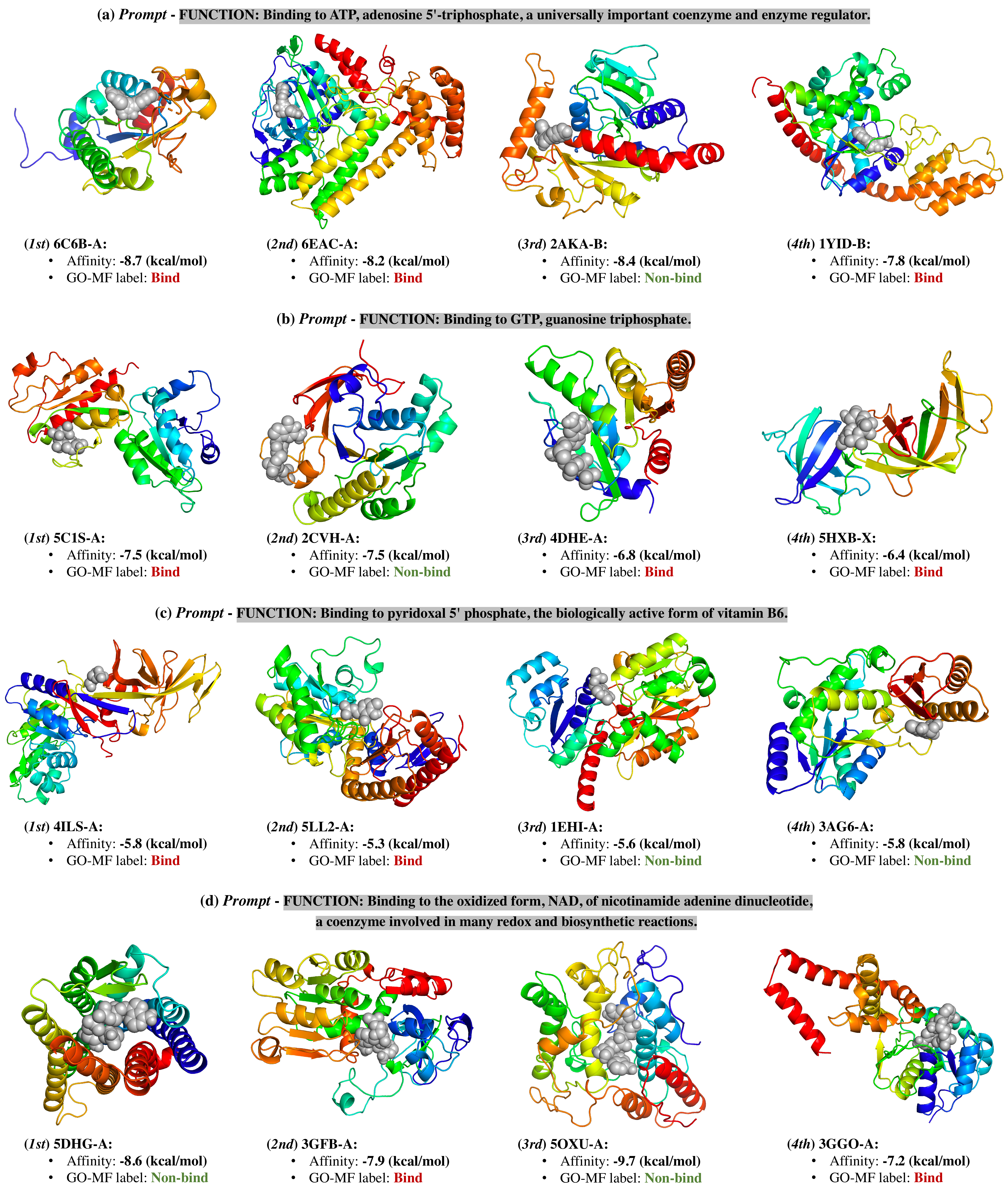}
    \vspace{-6mm}
    \caption{Zero-shot text-to-protein retrieval of (a) ATP binders, (b) GTP binders, (c) P5P binders, and (d) NAD+ binders based on {\method}-ESM-1b.}
    \label{supp:fig:t2p}
\end{figure*}

In Fig.~\ref{supp:fig:t2p}, we study four more sets of text-to-protein retrieval of ligand binders based on {\method}-ESM-1b. For each study, we visualize the text prompt and the top-4 retrieved candidates. For each candidate, we present the docking result of it binding with the ligand, the binding affinity and its GO molecular function label of binding with the ligand, where AutoDock Vina~\cite{trott2010autodock} is used to estimate docking pose and binding affinity. It is observed that, among the top-4 candidates, {\method}-ESM-1b succeeds in retrieving 3 GO-annotated ATP binders (only 3.99\% proteins are annotated as ATP binders in GO), 3 GO-annotated GTP binders (only 1.18\% proteins are annotated as GTP binders in GO), 2 GO-annotated P5P binders (only 0.17\% proteins are annotated as P5P binders in GO), and 2 GO-annotated NAD+ binders (only 0.05\% proteins are annotated as NAD+ binders in GO). The rest candidates annotated as non-binding also own decent binding affinity, \emph{e.g.}, the better binding affinity of protein 2AKA-B (\emph{without} ATP binder annotation) against protein 6EAC-A (\emph{with} ATP binder annotation), the better binding affinity of protein 5DHG-A (\emph{without} NAD+ binder annotation) against protein 3GFB-A (\emph{with} NAD+ binder annotation), \emph{etc.} These results demonstrate the general effectiveness of {\method}-ESM-1b on retrieving the binders of diverse ligands. In the future work, we will study how {\method} enables zero-shot text-to-protein retrieval of other types of functional proteins, \emph{e.g.}, antigen binders, toxic substance binders, transcription factors, \emph{etc.} 

%%%%%%%%%%%%%%%%%%%%%%%%%%%%%%%%%%%%%%%%%%%%%%%%%%%%%%%%%%%%

%%%%%%%%%%%%%%%%%%%%%%%%%%%%%%%%%%%%%%%%%%%%%%%%%%%%%%%%%%%%

\section{More Ablation Study}  \label{supp:sec:ablation}

%%%%%%%%%%%%%%%%%%%%%%%%%%%%%%%%%%%%%%%%%%%%%%%%%%%%%%%%%%%%

\begin{table}[t]
\begin{spacing}{1.05}
\centering
\caption{\small Ablation study of BLM on localization and fitness prediction. {\method}-ESM-1b serves as the base model. \emph{Abbr.}, Loc.: Localization; pred.: prediction; Acc: accuracy.}
\vspace{-2.5mm}
\label{supp:tab:ablation:blm:loc-fit}
\begin{adjustbox}{max width=1\linewidth}
% \begin{threeparttable}
    \begin{tabular}{l|cc|ccccc|c}
        \toprule
        \multirow{2}{*}{\bf{BLM}} & \multicolumn{2}{c|}{\bf{Loc. pred.} (\emph{Acc\%})} &
        \multicolumn{6}{c}{\bf{Fitness pred.} (\emph{Spearman's \bm{$\rho$}})} \\
        \cmidrule{2-3}
        \cmidrule{4-9}
        & \bf{Bin} & \bf{Sub} & \bf{$\beta$-lac} & \bf{AAV} & \bf{Thermo} & \bf{Flu} & \bf{Sta} & \bf{Mean \bm{$\rho$}} \\
        \midrule
        \multicolumn{9}{c}{\bf{Fix-encoder learning}} \\
        \midrule
        PubMedBERT-abs & 92.87 & 82.00 & \textbf{0.578} & \textbf{0.460} & 0.680 & \textbf{0.523} & \textbf{0.766} & \textbf{0.601} \\
        PubMedBERT-full & \textbf{93.04} & \textbf{82.28} & 0.548 & 0.458 & \textbf{0.682} & 0.507 & 0.744 & 0.588 \\
        \midrule
        \multicolumn{9}{c}{\bf{Full-model tuning}} \\
        \midrule
        PubMedBERT-abs & 92.35 & 78.73 & 0.895 & \textbf{0.850} & \textbf{0.681} & \textbf{0.682} & \textbf{0.751} & \textbf{0.772} \\
        PubMedBERT-full & \textbf{92.87} & \textbf{78.77} & \textbf{0.899} & 0.785 & 0.672 & 0.680 & 0.722 & 0.752 \\
        \bottomrule
    \end{tabular}
% \begin{tablenotes}
%     \item[*] Fix-encoder learning: the PLM is used as a feature extractor with pre-trained weights frozen.
% \end{tablenotes}
% \end{threeparttable}
\end{adjustbox}
\end{spacing}
\vspace{-2mm}
\end{table}

%%%%%%%%%%%%%%%%%%%%%%%%%%%%%%%%%%%%%%%%%%%%%%%%%%%%%%%%%%%%

\begin{table}[t]
\begin{spacing}{1.07}
\centering
\caption{\small Ablation study of BLM on function annotation. {\method}-ESM-1b serves as the base model.}
\vspace{-2.5mm}
\label{supp:tab:ablation:blm:anno}
\begin{adjustbox}{max width=1\linewidth}
    \begin{tabular}{l|cc|cc|cc|cc}
        \toprule
        \multirow{2}{*}{\bf{BLM}} & \multicolumn{2}{c|}{\bf{EC}} & \multicolumn{2}{c|}{\bf{GO-BP}} & \multicolumn{2}{c|}{\bf{GO-MF}} & \multicolumn{2}{c}{\bf{GO-CC}} \\
        \cmidrule{2-3}
        \cmidrule{4-5}
        \cmidrule{6-7}
        \cmidrule{8-9}
        & AUPR & $\mathrm{F}_{\mathrm{max}}$ & AUPR & $\mathrm{F}_{\mathrm{max}}$ & AUPR & $\mathrm{F}_{\mathrm{max}}$ & AUPR & $\mathrm{F}_{\mathrm{max}}$ \\
        \midrule
        \multicolumn{9}{c}{\bf{Full-model tuning}} \\
        \midrule
        PubMedBERT-abs & 0.894 & \textbf{0.878} & \textbf{0.328} & \textbf{0.480} & \textbf{0.644} & \textbf{0.661} & 0.364 & \textbf{0.488} \\
        PubMedBERT-full & \textbf{0.905} & \textbf{0.878} & 0.323 & 0.475 & 0.630 & 0.652 & \textbf{0.374} & 0.485 \\
        \bottomrule
    \end{tabular}
\end{adjustbox}
\end{spacing}
\end{table}

%%%%%%%%%%%%%%%%%%%%%%%%%%%%%%%%%%%%%%%%%%%%%%%%%%%%%%%%%%%%

\subsection{Ablation Study of Pre-training Losses} \label{supp:sec:ablation:loss}

In Tabs.~\ref{supp:tab:ablation:loss:loc-fit} and \ref{supp:tab:ablation:loss:anno}, we report the performance of {\method}-ESM-1b on all benchmark tasks by using full or partial pre-training losses. It can be observed that: (1) removing the loss $\mathcal{L}_{\mathrm{MPM}}$ leads to performance decay on 16 out of 24 benchmark metrics; (2) removing the loss $\mathcal{L}_{\mathrm{GC}}$ leads to decay on 20 out of 24 benchmark metrics; (3) removing the loss $\mathcal{L}_{\mathrm{MMP}}$ diminishes model performance on 19 out of 24 benchmark metrics. Therefore, all pre-training losses are necessary to maximize the effectiveness of a {\method}-induced PLM, where $\mathcal{L}_{\mathrm{GC}}$ and $\mathcal{L}_{\mathrm{MMP}}$ inject different granularities of protein property information into a PLM, and $\mathcal{L}_{\mathrm{MPM}}$ preserves the PLM's original representation power. 

%%%%%%%%%%%%%%%%%%%%%%%%%%%%%%%%%%%%%%%%%%%%%%%%%%%%%%%%%%%%

\begin{figure}[t]
\centering
\includegraphics[width=0.76\linewidth]{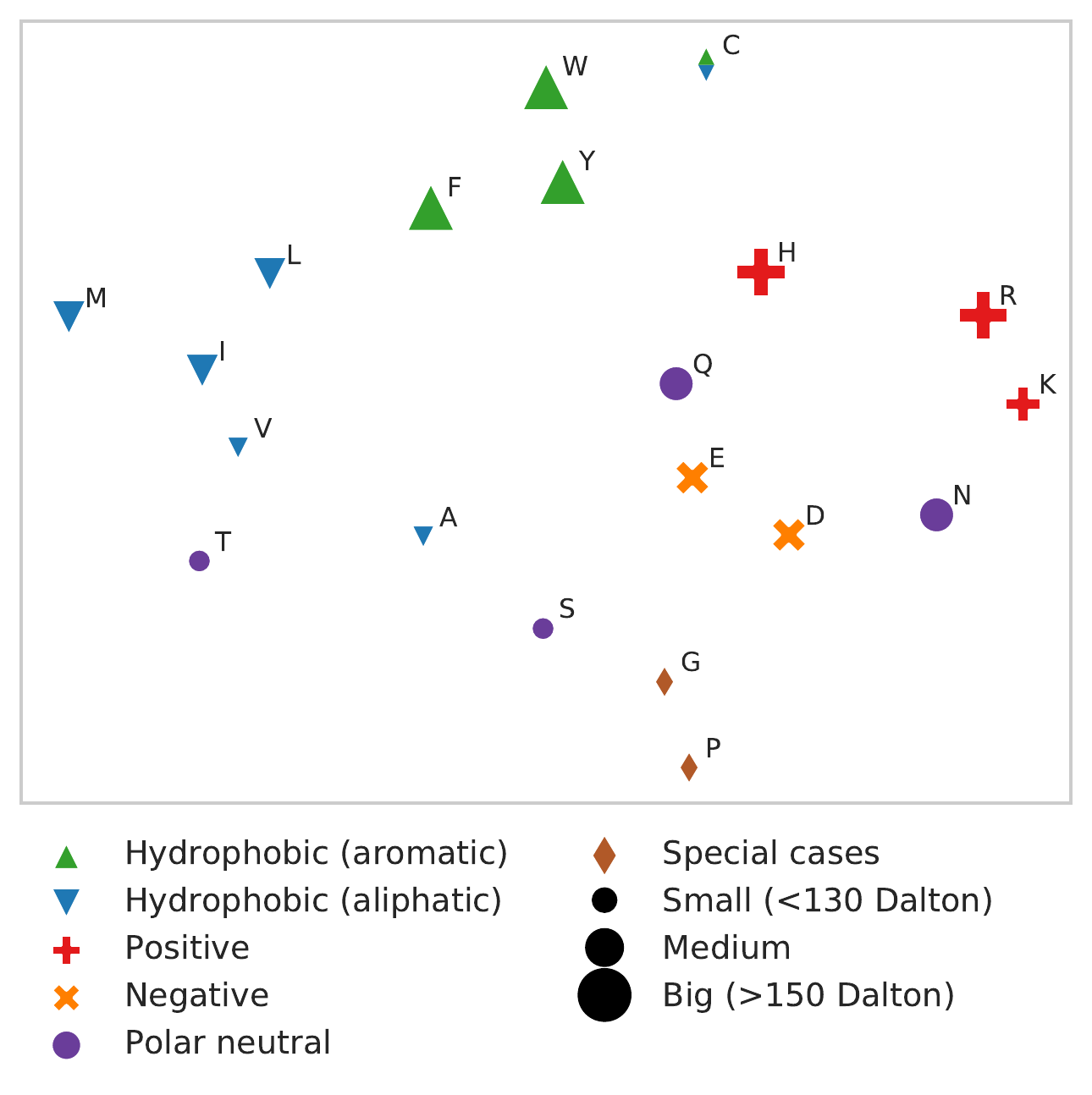}
\vspace{-3.8mm}
\caption{\small Amino acid representations learned by the linear layer for unimodal mask prediction ({\method}-ESM-1b is used).}
\label{fig:supp_tsne_aa_mlm}
\vspace{1.8mm}
\includegraphics[width=0.76\linewidth]{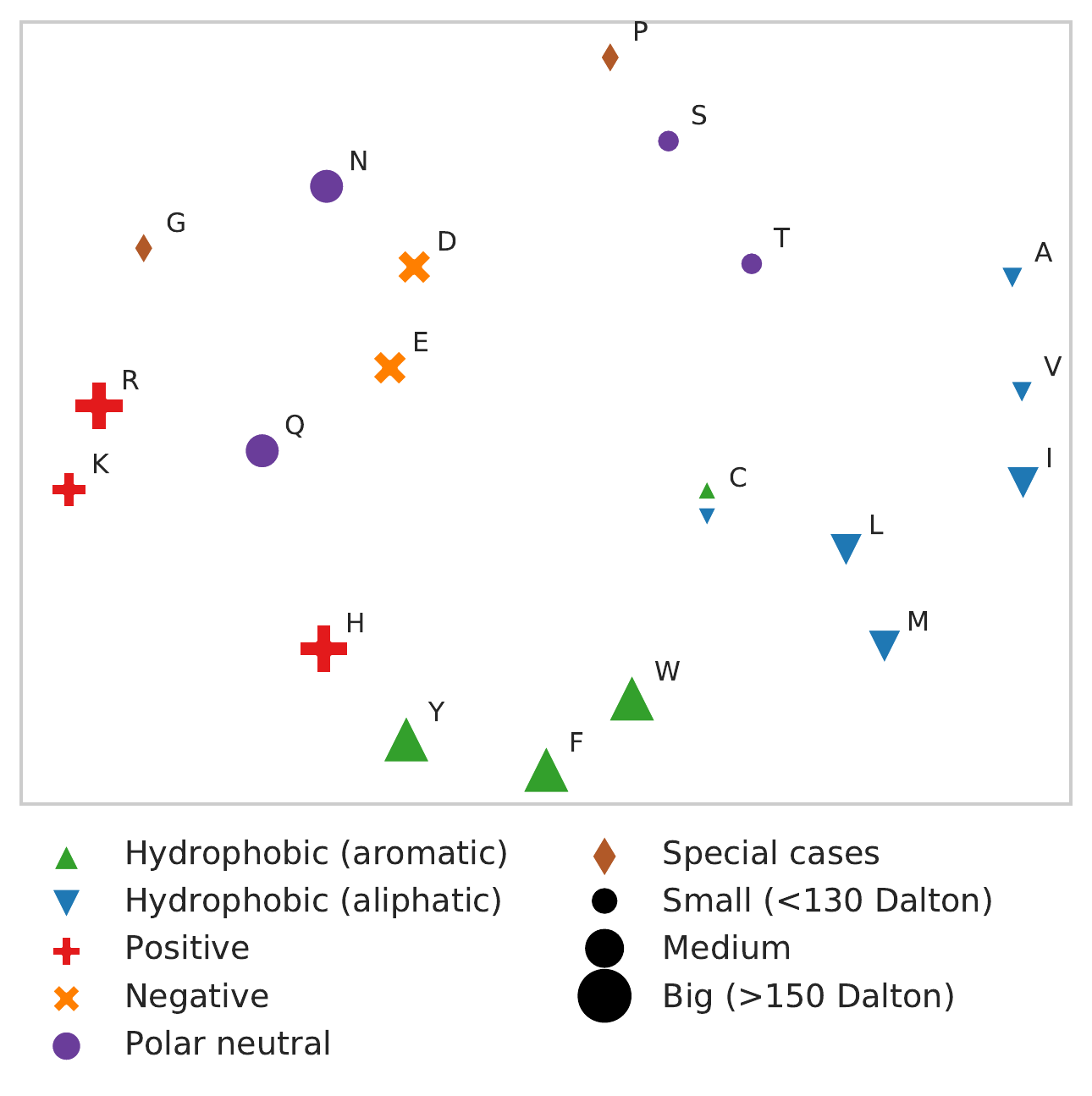}
\vspace{-3.8mm}
\caption{\small Amino acid representations learned by the linear layer for multimodal mask prediction ({\method}-ESM-1b is used).} 
\label{fig:supp_tsne_aa_mmp}
\vspace{-3mm}
\end{figure}

%%%%%%%%%%%%%%%%%%%%%%%%%%%%%%%%%%%%%%%%%%%%%%%%%%%%%%%%%%%%

\subsection{Ablation Study of Biomedical Language Model} \label{supp:sec:ablation:blm}

PubMedBERT owns two versions: (1) the PubMedBERT-abs trained by using only PubMed abstracts, and (2) the PubMedBERT-full trained by using additional PubMed Central full-text articles. In this experiment, we compare the effectiveness of these two models by respectively using them as the BLM of {\method}-ESM-1b. 

Tabs.~\ref{supp:tab:ablation:blm:loc-fit} and \ref{supp:tab:ablation:blm:anno} report the performance comparison of these two models on all benchmark tasks. We can observe that: (1) PubMedBERT-full outperforms PubMedBERT-abs on all four benchmark metrics of localization prediction; (2) PubMedBERT-abs performs better than PubMedBERT-full on 10 out of 12 benchmark metrics of fitness prediction; (3) PubMedBERT-abs outperforms PubMedBERT-full on 5 out of 8 benchmark metrics of function annotation. Therefore, PubMedBERT-full does not show superiority over PubMedBERT-abs in {\method} pre-training, which owes to the fact that the protein property descriptions in the {\dataset} dataset are more like abstracts than full-text articles. 

%%%%%%%%%%%%%%%%%%%%%%%%%%%%%%%%%%%%%%%%%%%%%%%%%%%%%%%%%%%%

%%%%%%%%%%%%%%%%%%%%%%%%%%%%%%%%%%%%%%%%%%%%%%%%%%%%%%%%%%%%

\section{More Visualization}  \label{supp:sec:visualize}

%%%%%%%%%%%%%%%%%%%%%%%%%%%%%%%%%%%%%%%%%%%%%%%%%%%%%%%%%%%%

\begin{figure}[t]
\centering
\includegraphics[width=0.85\linewidth]{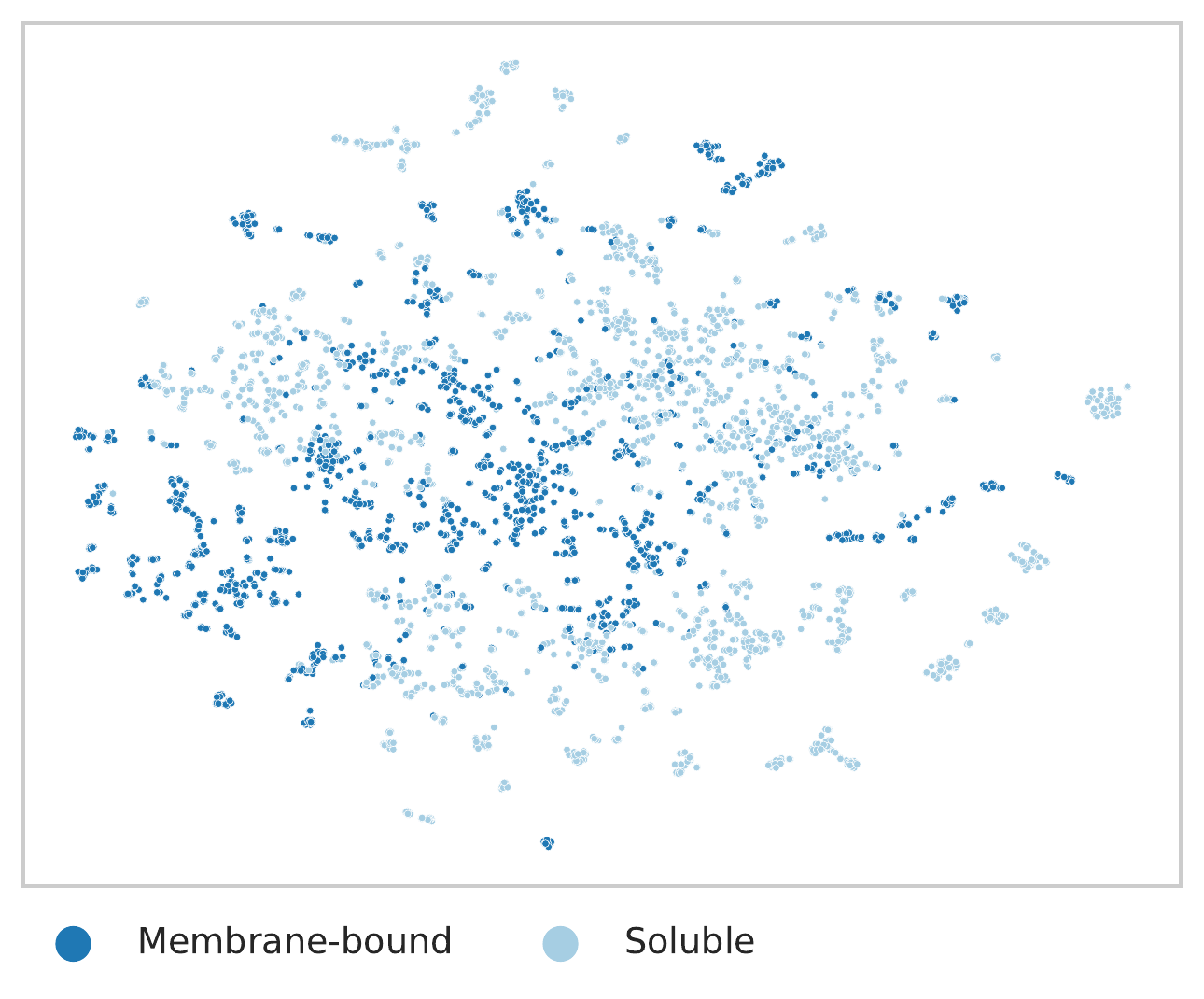}
\vspace{-3.8mm}
\caption{\small Visualization of protein representations on the binary localization prediction dataset ({\method}-ESM-1b is used).}
\label{fig:supp_tsne_binloc}
\vspace{1.8mm}
\includegraphics[width=0.85\linewidth]{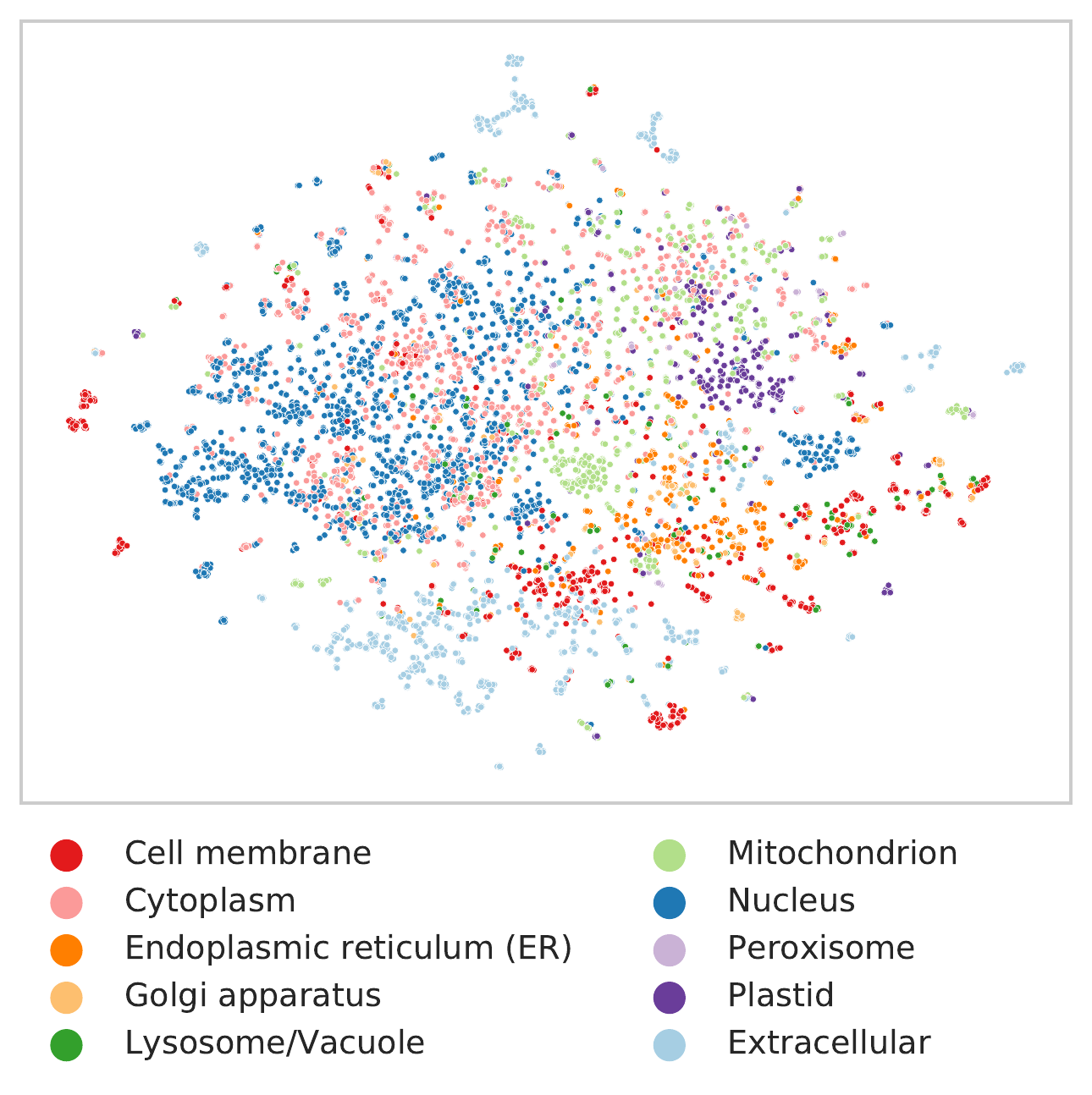}
\vspace{-3.8mm}
\caption{\small Visualization of protein representations on the subcellular localization prediction dataset ({\method}-ESM-1b is used).}
\label{fig:supp_tsne_subloc}
\vspace{-3mm}
\end{figure}

%%%%%%%%%%%%%%%%%%%%%%%%%%%%%%%%%%%%%%%%%%%%%%%%%%%%%%%%%%%%

Well-trained PLMs should have the capacity to extract structural, functional, and even evolutionary features
of proteins. As a result, the learned representations in PLMs are expected to have certain intrinsic organization patterns in the embedding space to capture these protein characteristics. To demonstrate the effectiveness of {\method}-ESM-1b, we use t-SNE \cite{van2008visualizing} to visualize such information  at different scales from amino acid decompositions to protein functional properties.

\textbf{Biophysical Properties of Amino Acids:} 
It is known that the biophysical properties of amino acids, such as hydrophobicity, aromaticity and charge, highly influence the biological structures of proteins and therefore their biological functions as well. To investigate if {\method}-ESM-1b captures such intrinsic features, we apply t-SNE to the two linear layers
% weight matrices of the final token embedding layer 
used for unimodal mask prediction and multimodal mask prediction. As shown in Figs.~\ref{fig:supp_tsne_aa_mlm} and \ref{fig:supp_tsne_aa_mmp}, hydrophobic and polar residues exhibit clear distinct clusterings, even to the level of aliphatic \emph{v.s.} aromatic. The clustering is also coherent in terms of the charge and size of the amino acids.

\textbf{Biological and Biochemical Properties of Proteins:} As introduced in Sec.~\ref{sec:exp:pretrain},  our proposed {\dataset} dataset provides {\method}-ESM-1b with direct access to knowledge like protein subcellular localizations, which refers to a specific region within a cell where the proteins can be found. For a protein, such locations can influence its activity and interaction with other molecules, thus helping the PLMs to better capture the biological and biomedical protein functions. To validate this assumption, we adopt the datasets used in two protein localization prediction tasks, \emph{i.e.}, the subcellular localization prediction and the binary localization prediction. With t-SNE, we project protein representations 
% in the protein-text-aligned space 
to the 2-dimensional space for these two benchmark datasets. In Figs.~\ref{fig:supp_tsne_binloc} and \ref{fig:supp_tsne_subloc}, certain clustering patterns of different cellular locations are observed. 
% as anticipated.

%%%%%%%%%%%%%%%%%%%%%%%%%%%%%%%%%%%%%%%%%%%%%%%%%%%%%%%%%%%%

%%%%%%%%%%%%%%%%%%%%%%%%%%%%%%%%%%%%%%%%%%%%%%%%%%%%%%%%%%%%

%%%%%%%%%%%%%%%%%%%%%%%%%%%%%%%%%%%%%%%%%%%%%%%%%%%%%%%%%%%%%%%%%%%%%%%%%%%%%%%
%%%%%%%%%%%%%%%%%%%%%%%%%%%%%%%%%%%%%%%%%%%%%%%%%%%%%%%%%%%%%%%%%%%%%%%%%%%%%%%

\end{document}